\documentclass[fleqn,usenatbib]{mnras}

\usepackage{newtxtext,newtxmath}
\usepackage{natbib}

\usepackage[toc,page]{appendix}
\usepackage{rotating}
\usepackage{mathtools}
\usepackage{tablefootnote}
\usepackage[para]{threeparttable}
\usepackage{tabularx}
\usepackage{placeins}
\usepackage{multirow}
\usepackage{graphicx}
\usepackage{url}
\usepackage{adjustbox}
\usepackage{float}

\usepackage{soul}

\usepackage[usenames,dvipsnames,svgnames]{xcolor}

\newcommand{\Mmax}{M_{\rm bry}^{\rm max}}
\newcommand{\Erot}{\mathcal{T}}
\newcommand{\Msol}{\mathrm{M}_\odot}

\usepackage[T1]{fontenc}

\usepackage{graphicx}	
\usepackage{amsmath}	

\title[Nucleosynthesis in magneto-rotational supernovae]{Nucleosynthesis in magneto-rotational supernovae}

\author[M. Reichert et al.]{
M. Reichert,$^{1,2}$
M. Obergaulinger,$^{1,3}$
M. Eichler,$^{1}$
M. \'{A}. Aloy,$^{3}$
and A. Arcones$^{1,2,4}$
\thanks{E-mails: \newline mreichert@theorie.ikp.physik.tu-darmstadt.de \newline miguel.a.aloy@uv.es \newline almudena.arcones@physik.tu-darmstadt.de}
\\
$^{1}$Institut f\"ur Kernphysik, Technische Universit\"at Darmstadt, Schlossgartenstr. 2,
Darmstadt 64289, Germany\\
$^{2}$Helmholtz Forschungsakademie Hessen f\"ur FAIR, GSI Helmholtzzentrum für
Schwerionenforschung, 64291 Darmstadt, Germany\\
$^{3}$Departament d'Astonomia i Astrof\'{\i}sca, Universitat de Val\`encia, Edifici d'Investigatci\'{o} Jeroni Munyoz, C/Dr. Moliner, 50, E-46100 Burjassot (Val\`encia), Spain\\
$^4$GSI Helmholtzzentrum f\"ur Schwerionenforschung GmbH, Planckstr. 1, Darmstadt 64291, Germany
›} 

\date{Accepted XXX. Received YYY; in original form ZZZ}

\pubyear{2020}

\begin{document}
\label{firstpage}
\pagerange{\pageref{firstpage}--\pageref{lastpage}}
\maketitle

\begin{abstract}
We present the nucleosynthesis of magneto-rotational supernovae (MR-SNe) including neutrino-driven and magneto-rotational-driven ejecta based, for the first time, on 2D simulations with accurate neutrino transport. The models analysed here have different rotation and magnetic fields, allowing us to explore the impact of these two key ingredients. The accurate neutrino transport of the simulations is critical to analyse the slightly neutron-rich and proton-rich ejecta that are similar to the, also neutrino-driven, ejecta in standard supernovae. In the model with strong magnetic field, the r-process produces heavy elements up to the third r-process peak ($A\sim 195$), in agreement with previous works. This model presents a jet-like explosion with proton-rich jets surrounded by neutron-rich material where the r-process occurs. We have estimated a lower limit for $^{56}$Ni of $2.5\times10^{-2} M_\odot$, which is still well below the expected hypernova value. Longer simulations including the accretion disc evolution are required to get a final prediction. In addition, we have found that the late evolution is critical in a model with weak magnetic field in which late-ejected neutron-rich matter produces elements up to the second r-process peak. Even if we cannot yet provide conclusions for hypernova nucleosynthesis, our results agree with observations of old stars and radioactive isotopes in supernova remnants. This makes MR-SNe a good additional scenario to neutron star mergers for the synthesis of heavy elements and brings us closer to understand their origin and the role of MR-SNe in the early Galaxy nucleosynthesis.
\end{abstract}


\section{Introduction}
\label{sec:intro}
Core-collapse supernovae are critical for the chemical history of the universe. These explosive events at the end of the life of massive stars enrich the interstellar medium with alpha elements, iron group elements, and probably heavier ones. What are the heaviest
elements that can be produced in core-collapse supernovae? This
depends on how the matter is ejected. Standard supernovae are driven by
neutrinos \citep[see e.g., ][for recent reviews]{Janka2016, Mueller2020,Kotake2012} that determine the conditions of the ejecta, namely electron
fraction ($Y_e$), entropy, and expansion time-scale. The ejecta can be
slightly neutron-rich ($Y_e<0.5$) and/or proton rich ($Y_e>0.5$).
In both cases, elements up to Sr, Y, Zr may be produced
\citep{Arcones2013,ArconesBliss2014,Wanajo2018}. Alternatives to this standard mechanism
have been suggested to explain very energetic supernovae, some of
which are associated with long gamma-ray bursts (GRBs) producing
relativistic outflows \citep{Nomoto2006,Woosley_Bloom__2006__araa__The_Supernova_Gamma-Ray_Burst_Connection,Cano_et_al__2017__AdvancesinAstronomy__TheObserversGuidetotheGamma-RayBurstSupernovaConnection,Moriya_et_al__2018__ssr__SuperluminousSupernovae}.  Since by themselves, neutrino heating and
hydrodynamic instabilities have difficulties powering these extreme
events, rapid rotation and strong magnetic fields have been invoked to
explain these events \citep{Wheeler_et_al__2002__apj__Asymmetric_SNe_from_Magnetocentrifugal_Jets,Maeda_Nomoto__2003__apj__Bipolar_SN_Nucleosynthesis_and_Implications_for_Abundances_in_EMP_Stars,Dessart_et_al__2008__apjl__TheProto-NeutronStarPhaseoftheCollapsarModelandtheRoutetoLong-SoftGamma-RayBurstsandHypernovae,Tominaga__2009__apj__Aspherical_Properties_of_Hydrodynamics_and_Nucleosynthesis_in_Jet-Induced_Supernovae,Metzger_et_al__2011__mnras__Theprotomagnetarmodelforgamma-raybursts,Dessart_et_al__2012__mnras__Superluminoussupernovae:$56$Nipowerversusmagnetarradiation,Mazzali_et_al__2014__mnras__Anupperlimittotheenergyofgamma-rayburstsindicatesthatGRBs/SNearepoweredbymagnetars,Metzger_et_al__2015__mnras__Thediversityoftransientsfrommagnetarbirthincorecollapsesupernovae,Metzger_etal_2018ApJ...857...95,Aloy_Obergaulinger_PaperII_2020arXiv200803779A}.  Further indications for the importance of
magnetic fields in a subset of all supernovae come from the
observation of very strongly magnetized, young neutron stars,
so-called magnetars \citep{Duncan1992,Kaspi_Beloborodov__2017__AnnualReviewofAstronomyandAstrophysics__Magnetars}.  If combined with high rotational energies, newly
born magnetars (also known as \emph{protomagnetars}) could inject energy at high rates into the ejecta and
power very violent explosions, thereby spinning down to their
observed, rather long rotation periods \citep{Metzger_et_al__2011__mnras__Theprotomagnetarmodelforgamma-raybursts}. This explosion mechanism has
not attracted the same attention as standard explosions (but see \citealt{Aloy_Obergaulinger_PaperII_2020arXiv200803779A}, who show evidences of protomagnetar formation more than 5\,s after the core bounce of low-metallicity, massive, stellar progenitors endowed with sufficiently strong poloidal magnetic fields). Partly, this is due to
numerical difficulties such as the necessity to resolve small-scale
structures of the flow and the field generated by, e.g., the
magnetorotational instability \citep{Obergaulinger2009,Rembiasz_2016MNRAS.456.3782,Rembiasz_2016MNRAS.460.3316}.  
Additionally, the required rotation rates and magnetic energies restrict this mechanism to a minority of progenitor stars and, thus, observed explosions. 

Despite their relatively small numbers, magneto-rotational supernovae
(MR-SNe) may nevertheless be important contributors to the enrichment
of galaxies with heavy elements in the early universe \citep{Cote2019}. In addition to neutrino-driven ejecta, these explosions have an early and fast ejection of matter where the 
rapid neutron capture process (r-process) can efficiently produce
heavy elements \citep{Nishimura2006,Winteler2012,Saruwatari2013,Nishimura2015,Nishimura2017,Moesta2018}, similarly to the prompt
explosions found in the 70s \citep{Hillebrandt1976}. Here we present the first nucleosynthesis
study based on 2D supernova simulations with accurate
neutrino transport \citep{ObergaulingerAloy2017,ObergaulingerAloy2020,Aloy_Obergaulinger_PaperII_2020arXiv200803779A}. Therefore, we can uniquely and
consistently study the nucleosynthesis of both neutrino-driven and
magnetic-driven ejecta within the same supernova model.

The nucleosynthesis in core-collapse supernovae have been extensively
studied for alpha and iron group elements based on simple models
\cite[see e.g., ][for thermal bombs, piston, parametric neutrino
heating explosions]{Woosley1995,Thielemann1996,Rauscher2002,Nomoto2006,Woosley2007,Umeda2008,Heger2010,Nomoto2013,Sukhbold2016,Chieffi2017,Nomoto2017,Limongi2018,Curtis2019,Ebinger2020,Ertl_2020ApJ...890...51} with parameters
fixed to reproduce observations of explosion energy and Ni yields. In
addition, supernovae were suggested as the r-process sites where half
of the elements beyond iron are produced \citep{Burbidge1957,Woosley1994}. With the improvement of simulations and neutrino
treatment, it became clear that standard supernovae cannot produce
elements beyond the second r-process peak, $A\sim 130$ \cite[see][for
a review and references within]{Arcones2013}. Current
simulations show that the ejecta is often proton rich with some small,
fast-expanding clumps of slightly neutron-rich material \citep{Wanajo2011,Obergaulinger_Aloy_2020__mnras_PaperIII}. Therefore, elements between the iron group and the second r-process
peak may be synthesized by a weak r-process under slightly
neutron-rich conditions \cite{Wanajo2011, ArconesBliss2014, Bliss2020}  and/or by the $\nu$p-process in
proton-rich conditions \citep{Froehlich2006, Pruet2006, Wanajo2006}. In addition, observations of heavy
r-process elements at low metallicities present a large scatter as a function of metallicity compared to iron group and alpha elements produced in core-collapse supernovae \citep[see e.g.,][for recent reviews]{Sneden2008,Cowan2019}. This suggests that the r-process occurs only in rare events, and they may not happen in every core-collapse supernova. In the 1970s, the merger of a black hole and a neutron star was suggested as possible rare r-process event \citep{Lattimer1974}. In 2017, the production of r-process elements in neutron star mergers was confirmed by the observation of the kilonova light curve triggered by radioactive decay of neutron-rich nuclei after the gravitational wave detection of the merger GW170817 \citep{Abbott_2017ApJ...848L..12_MultimessengerGW170817, Abbott2017,Kasen_2017Natur.551...80,Pian_2017Natur.551...67_Spectroscopic_identification_of_r-process,Smartt2017,Tanvir_2017ApJ...848L..27}. However, such events have difficulties to explain the heavy elements observed in the oldest stars and the trend of Eu-over-Fe abundance ratios with metallicity \citep{Cote2019}. An additional r-process site has to exist at low metallicities and MR-SNe are a promising candidate.

 \cite{LeBlanc1970} and later \cite{Cameron2003} proposed MR-SNe as r-process site and \citet{Nishimura2006} presented the first successful results based on 2D adiabatic MHD simulations (i.e., ignoring the effects of neutrino cooling and heating). Similar results
were found also with 3D simulations by \citet{Winteler2012}, using a simplified and post-processed prescription of neutrino heating. For the rotation and magnetic fields, there are still many uncertainties in the progenitor models \citep[e.g.][]{Meynet_2018IAUS..334..170}. Based on simple neutrino treatment, \citet{Nishimura2015} have investigated the impact of different magnetic field strengths, and rotation rates on the nucleosynthesis. \citet{Nishimura2017} further improved the previous study by considering the effects of the magnetic field enhancement as a result of the magnetorotational instability. Moreover, they also vary neutrino luminosities, which are not fully consistent but only a parameter in their models. In addition,  
\citet{Moesta2018} showed that the assumption of
2D may artificially support the production of heavy elements
and that their 3D models needed even stronger magnetic fields to successfully
produce heavy elements. They assumed that the neutron-rich material and thus the r-process occurs in the collimated ejecta and argue that this jet-like structure is not a robust 3D feature due to the kink instabilities. Also, a misalignment of the magnetic field
with respect to the rotational axis can have an influence on
the neutron-richness of the ejecta (reducing it), so that the r-process becomes weaker \citep{Halevi2018}. For the neutrinos, there are less uncertainties than for the magnetic field, but only recently it has been possible to perform MHD simulations with accurate neutrino transport, first in 2D \citep{Obergaulinger2014,ObergaulingerAloy2017,Aloy_Obergaulinger_PaperII_2020arXiv200803779A} and recently in 3D \citep{Obergaulinger_Aloy_2020__mnras_PaperIII,ObergaulingerAloy2020, Kuroda2020}.

Other potential r-process sites associated with MR-SNe are the accretion
discs that form after the explosion surrounding a massive neutron star
(magnetars) or a black hole (collapsars). Pioneering nucleosynthesis
studies \citep{Surman2004, McLaughlin2005,Surman2006} have demonstrated that neutrinos will
play a critical role reducing the neutron-richness of the ejecta and
thus the possibilities for the r-process. Recent works are not conclusive \citep{Miller2019,Siegel2019} and more work is required to understand the nucleosynthesis from supernova accretion discs.

In this paper, we have investigated the early explosive
nucleosynthesis in MR-SNe based on the first 2D simulations
that include a detailed neutrino transport treatment
\citep{ObergaulingerAloy2017,Obergaulinger_Aloy_2020__mnras_PaperIII,ObergaulingerAloy2020,Aloy_Obergaulinger_PaperII_2020arXiv200803779A}. Advancing beyond state-of-the-art \citep[see e.g., ][]{Nishimura2006,Winteler2012,Nishimura2015,Nishimura2017,Moesta2018}, consisting on parametrizing rotation, magnetic field, and neutrinos, here we employ a self-consistent neutrino treatment. For the rotation and magnetic field, we start with the predictions from stellar evolution \citep{woosley-heger2006, ObergaulingerAloy2017,ObergaulingerAloy2020} and vary them within the uncertainties that may result from stellar evolution and its mapping to multidimensional initial models for magnetorotational core collapse \citep{Aloy_Obergaulinger_PaperII_2020arXiv200803779A}. In addition, to the original progenitor values, we use simulations with increased and decreased magnetic field and also increased rotation. A total of four models are analysed and found that the r-process can occur only in the model with moderately enhanced, topologically dipolar magnetic field. This explosion develops jets that become proton rich. Very neutron-rich matter is only promptly ejected and stays around the jet. We have also found that, during the late evolution (more than 1 s after bounce), the angular momentum
redistribution can lead to a late ejection of  neutron-rich material. In this model, we find a weak r-process producing elements up to the second r-process peak ($A\sim 130$).

This paper is structured as follows. In Sec.~\ref{sec:method}, we
describe the magnetohydrodynamic models and the nuclear reaction
network. The nucleosynthesis and dynamics of the ejecta are presented
in Sec.~\ref{sec:results}. In Sec.~\ref{sec:obs}, we briefly compare our
results to different observations. Summary and conclusions are in Sec.~\ref{sec:con}.


\section{Methods}
\label{sec:method}

\subsection{MHD simulations: code and input physics}
\label{sec:MHDcode}

We calculate the nucleosynthesis of four of the models whose dynamics has been described by \cite{ObergaulingerAloy2017, ObergaulingerAloy2020}. The simulations of the collapse and the explosion of the stellar cores were performed using the radiation-magnetohydrodynamics (MHD) code {\sc aenus-alcar} \citep{Just2015}. The dynamics of the gas and the magnetic field were modelled using the equations of special relativistic MHD.  At high densities, $\rho > 6 \times 10^{7} \, \mathrm{g \, cm^{-3}}$, the system is closed by the equation of state (EOS) of \cite{LattimerSwesty1991} with an incompressibility of $K = 220 \, \mathrm{MeV}$. The EOS is tabulated on a 3D grid of density, temperature, and electron fraction.  The range in $Y_e$ is limited to a maximum of $Y_{e;\mathrm{max}} = 0.56$.  We encounter regions of our models in which the gas exceeds this maximum value and where we therefore rely on an extrapolation of the EOS in order to determine the thermodynamics of the gas.  Below the density threshold, we use an EOS based on a gas of electrons, positrons, photons, and baryons \citep{RamppJanka2002}. For the baryonic component, we used the so-called flashing scheme that assumes that matter is composed by a mixture of five nuclei, viz.~protons, neutrons, $\alpha$-particles, Si, and Ni nuclei. We accounted for the effects of general relativity in the gravitational field by using one of the post-Newtonian TOV potentials \citep[version 'A'; see][]{Obergaulinger_2006A&A...457..209} of \cite{Mareketal2006}.

Neutrinos are treated in the spectral two-moment, or M1, framework derived by expanding the Boltzmann equation of radiative transfer into angular moments of the phase-space distribution function of the neutrinos. This expansion yields balance equations for the energy and momentum densities of the neutrinos. The system is closed by a local algebraic Eddington tensor. We evolve the neutrino moments in the frame comoving with the gas and include energy-bin coupling terms involving the fluid velocity and gravitational potential in the $v/c$-plus approximation of \cite{Endeve2012}. Matter and neutrinos couple via the following reactions: emission and absorption of neutrinos by nucleons and nuclei, scattering of nucleons, nuclei, and electrons, electron-positron pair annihilation, and nucleonic bremsstrahlung.  For more details, see \cite{Obergaulinger_2014ASPC..488..255} and \cite{Obergaulinger_2018JPhG...45h4001}.

Our simulations are based on model 35OC for a star of an initial mass
$M_{\mathrm{ZAMS}} = 35 \, \mathrm{M}_\odot$ from the stellar-evolution
calculations by \citet{woosley-heger2006}.  Rotation and magnetic
fields were included in the spherically symmetric models following the
recipe of \citet{2002A&A...381..923S}. Within the series of four
models to which this progenitor belongs to, the mass-loss was a free parameter. As our reference model (35OC-RO), we selected the one with
the second smallest mass-loss, which at collapse has a mass of $M =
28.1 \, \mathrm{M}_\odot$ and  an iron core of 
$M_\mathrm{Fe} = 2.02 \, \mathrm{M}_\odot$.  It rotates differentially
with an angular velocity $\Omega_c = 1.98$\,Hz at the centre and $\Omega_\mathrm{Fe} \approx 0.1\,$Hz at the surface of the Fe core. The data contain the radial profiles of the poloidal and toroidal components of the magnetic field in the radiative layers. In convectively unstable layers, the field is set to zero by construction. With a field strength of $b^{\mathrm{pol};\mathrm{tor}} \approx 1.7 \times 10^{10} ; 1.7 \times 10^{11} \, \mathrm{G}$ for the poloidal and toroidal components at the centre of the core, the model possesses a relatively, though not extremely,  high magnetization. We constructed the 2D
distribution of the magnetic field from these radial profiles by assuming a
sine dependence in $\theta$ \citep[see][]{Aloy_Obergaulinger_PaperII_2020arXiv200803779A}. The spherical grid used in our models consists of $(n_r,n_\theta)=(400,128)$ zones, uniform in the polar angle $\theta$ and unevenly distributed in the radial direction \citep[see][for details]{ObergaulingerAloy2020}.

\subsection{Supernova models}
\label{sec:models}
The nucleosynthesis presented here is based on the four supernova models.
We varied the original profiles of the rotational velocity and the magnetic field of model
35OC-RO to set-up the other three models. 
35OC-Rw and 35OC-Rs are based on the same rotational profile, but
replacing the magnetic field by an artificial distribution of poloidal
and toroidal field following the prescription of
\citet{2007PASJ...59..771S}. The normalisation of the field strengths
is $b^{\mathrm{pol}}=b^{\mathrm{tor}} = 10^{10} \,\mathrm{G}$ for 35OC-Rw and
$b^{\mathrm{pol}}=b^{\mathrm{tor}} = 10^{12} \,\mathrm{G}$ for 35OC-Rs, respectively. Model
35OC-RRw has an initial field that is six orders of magnitude weaker than that of 35OC-Rw and thus dynamically insignificant. It rotates $1.5$ times faster than model 35OC-Rw.

\begin{table*}
	\centering
	\caption{MR-SN models.}
	  \label{tab:models}
	\begin{tabular}{cccccccccc} 
		\hline
   Name  
    & Rotation$^{a }$ 
    & Magnetic field$^{b }$ 
    & $t_{\mathrm{fin}}$ $^{c }$ 
    & $t_{\mathrm{exp}}$ $^{d }$ 
    & $E_{\mathrm{exp}}$ $^{e }$ 
    & $M_{\mathrm{ej}}$ $^{f }$ 
    & $\mathcal{T}/|\mathcal{W}|$ $^{g }$ 
    & $\mathcal{B}/\mathcal{T}$ $^{h }$ 
    & type$^{i }$ 
    \\
    & & 
    & (s)
    & (ms)
    & (B) 
    & ($10^{-1}M_\odot$) 
    \\
    \hline
    35OC-RO & 1.0 & Or & 2.5 & 178 & 1.78 & 3.21 
    & 0.028 & 0.092
    & MR
    \\
    35OC-Rw & 1.0 & $10$ & 2.5 & 378 & 2.80 & 3.91 
    & 0.040 & 0.0089 & $\nu$-$\Omega$
    \\
    35OC-Rs & 1.0 & $12$ & 0.9 & 20 & 4.16 & 3.89 
    & 0.028 & 0.30 & MR
    \\
    35OC-RRw& 1.5 & $\mathrm{Or} \times 10^{-6}$ & 1.6 & 343 & 0.209 & 0.345 
    & 0.063 & $2.9 \times 10^{-5}$& $\nu$-$\Omega$
    \\
		\hline
\multicolumn{10}{p{\dimexpr\linewidth-2\tabcolsep-2\arrayrulewidth}}{$^a$ Increase of the pre-collapse rotational velocity w.r.t.~the original stellar evolution model.}\\
\multicolumn{10}{p{\dimexpr\linewidth-2\tabcolsep-2\arrayrulewidth}}{$^b$ Initial magnetic field: ``Or'' and ``$\mathrm{Or}\times 10^{-6}$'' denote the original field of the progenitor model ($b^{\mathrm{pol};\mathrm{tor}} \approx 1.7 \times 10^{10} ; 1.7 \times 10^{11} \, \mathrm{G}$) and the original field reduced by a uniform factor of $10^{-6}$, respectively, and a number $n$ indicates that the model was run using a normalization of both poloidal and toroidal components of $10^n \, \mathrm{G}$.}\\
\multicolumn{10}{p{\dimexpr\linewidth-2\tabcolsep-2\arrayrulewidth}}{$^c$ Time (post-bounce) of the last time-step of the simulations used for the nuclear network calculations (note that these models have been evolved for longer times in other publications, e.g.~\cite{ObergaulingerAloy2020} and \cite{Aloy_Obergaulinger_PaperII_2020arXiv200803779A}.}\\
\multicolumn{10}{p{\dimexpr\linewidth-2\tabcolsep-2\arrayrulewidth}}{$^d$ Time (post-bounce) at which an explosion is launched.}\\
\multicolumn{10}{p{\dimexpr\linewidth-2\tabcolsep-2\arrayrulewidth}}{$^e$ Diagnostic explosion energy at $t_{\mathrm{fin}}$.}\\
\multicolumn{10}{p{\dimexpr\linewidth-2\tabcolsep-2\arrayrulewidth}}{$^f$ Ejected mass at $t_{\mathrm{fin}}$.}\\
\multicolumn{10}{p{\dimexpr\linewidth-2\tabcolsep-2\arrayrulewidth}}{$^g$ Ratio of rotational to gravitational energy of the PNS at the time of explosion.}\\
\multicolumn{10}{p{\dimexpr\linewidth-2\tabcolsep-2\arrayrulewidth}}{$^h$ Ratio of magnetic to rotational energy of the PNS at the time of explosion.}\\
\multicolumn{10}{p{\dimexpr\linewidth-2\tabcolsep-2\arrayrulewidth}}{$^i$ "Type" gives a brief indication of the explosion type: $\nu$-$\Omega$ one strongly affected by rotation, MR a magnetorotational explosion.}\\
	\end{tabular}
\end{table*}

The four models evolve in fairly different ways (see Table~\ref{tab:models} for an overview of the models).  Model 35OC-Rw develops a neutrino-driven explosion after about $400 \; \mathrm{ms}$ post-bounce with a dynamically unimportant magnetic field at that time. Driven by strongly anisotropic neutrino fluxes, the explosion has the form of a relatively wide bipolar
outflow. The shock wave reaches a polar radius of
$R \approx 3 \times 10^4 \; \mathrm{km}$ at
$t \approx 2 \; \mathrm{s}$.  At that point, the ejecta contain an
energy of $\approx 6\times 10^{50}\; \mathrm{erg}$ and a mass of
$\approx 0.2 \, \mathrm{M}_\odot$.
The final values are $R \approx 4.7 \times 10^{4} \; \mathrm{km}$ for the maximum shock radius and $ 1.78 \times 10^{51} \; \mathrm{erg}$ and $\approx 0.321 \, \mathrm{M}_\odot$ for the ejecta energy and mass, respectively.
The proto-neutron star (PNS) grows in
mass by accretion to a baryonic mass of
$M \gtrsim \Mmax$, with $\Mmax=2.45\,\mathrm{M}_\odot$ being the maximum cold, non-rotating PNS mass for our EOS.  
It develops a high rotational energy of up to
$\Erot \approx 8\times 10^{52}\, \mathrm{erg}$.  The
magnetic energy in and around the PNS grows continuously, but
experiences a particularly strong increase after
$t \sim 1.8 \; \mathrm{s}$.  This growth causes a more efficient
redistribution of angular momentum from the central regions to the
outer layers of the PNS, where it is deposited \citep[see][section\,3.4]{Aloy_Obergaulinger_PaperII_2020arXiv200803779A}.  Consequently, growing centrifugal support leads to an
expansion of the PNS at low latitudes.  The increasingly oblate PNS
sheds mass from its equatorial regions and thus generates a surrounding neutron-rich layer, with important consequences
for the conditions of matter ejected.

 Model 35OC-RO explodes earlier ($t \approx 178 \; \mathrm{ms}$)
  due to the strong magnetic forces that play an important role in
  accelerating the outflows. They furthermore lead to a stronger
  collimation than in model 35OC-Rw. The explosion is faster and more
  energetic, reaching the radius of $3\times 10^{4}\; \mathrm{km}$ and
  an energy of $6\times 10^{50}\; \mathrm{erg}$ about $700 \;
  \mathrm{ms}$ earlier than 35OC-Rw, whereas the evolution of the
  ejecta mass is very similar in both cases. 
  By the end of the simulation, the shock expands to $R \approx 7.8 \times 10^{4} \; \mathrm{km}$  and the ejecta have an energy of $ 1.78 \times 10^{51} \; \mathrm{erg}$ and a mass of $\approx 0.321 \, \mathrm{M}_\odot$.
  The PNS is even more massive than in the previous model with 
  $M \approx 2.7\,\Msol$ at $t \approx 2 \; \mathrm{s}$.  Compared to
  model 35OC-Rw, the magnetic field in the PNS is stronger.  
  Hence, efficient angular-momentum redistribution and the
  associated high axis ratio of the PNS develop earlier.
  
 Model 35OC-Rs explodes almost immediately after core bounce. The
  explosion, driven entirely by the extremely strong magnetic fields,
  achieves a maximum radius of $3\times 10^{4}\; \mathrm{km}$ already at
  $t \approx 0.7 \; \mathrm{s}$. At that time, the energy and mass of
  the ejecta are $E \approx 3\times 10^{51} \mathrm{erg}$ and $0.35
  \, \mathrm{M}_\odot$, i.e., they grow faster than in any other model. The PNS
  on the other hand has the lowest mass because the strong explosion
  shuts down accretion and magnetic stresses even drive the lost of the outer PNS layers. It reaches $M \approx 1.9 \, \mathrm{M}_\odot$ at
  $t \approx 0.5 \;\mathrm{s}$ and afterwards tends to slowly lose
  mass. Without further accretion, the PNS also does not gain angular
  momentum, resulting in a comparably low rotational energy
  $\Erot \approx 2\times 10^{52}\, \mathrm{erg}$ at 
  $t \approx 0.7\; \mathrm{s}$.  Although the rotational energy is lower than in other models, 
  the axis ratio of the PNS exceeds theirs, because the extremely strong magnetic field accumulates
  a comparably large fraction of the angular momentum in the outer
  layers.

Model 35OC-RRw explodes at about the same time as 35OC-Rw,
  though less violently. At the end of the simulation ($t \approx 1.5
  \; \mathrm{s}$), the maximum shock radius is $R \approx 1.5\times
  10^{4}\; \mathrm{km}$, and the ejecta energy and mass are $E \approx
  2 \times 10^{50}\; \mathrm{erg}$ and $0.03 \, \mathrm{M}_\odot$,
  respectively, i.e., considerably less than 35OC-Rw at the same
  time. The reason for this weaker explosion is that the high
  rotational energy ($\Erot \approx 1.2\times 10^{53}\,\mathrm{erg}$ at $t = 1.5 \, \mathrm{s}$) reduces the accretion
  luminosity and, consequently, the neutrino heating rate. On the
  other hand, rotation allows for a high PNS mass of $M
  \approx 2.65 \; \Msol$ and an exceptionally high axis ratio. 
  A thorough overview of the post-bounce dynamics of all these models can be found in \cite{ObergaulingerAloy2017,ObergaulingerAloy2020} and \cite{Aloy_Obergaulinger_PaperII_2020arXiv200803779A}.


\subsection{Tracers and nucleosynthesis calculation}
\label{sec:netw}

The evolution of the ejecta is followed by Lagrangian tracer particles
that are set-up at the beginning of the simulations. Into each grid
cell, we insert four~tracer particles at random positions, corresponding
to a total number of 204 800 tracers in each model. Each particle
represents a fraction of one fourth of the total mass of the cell,
$m_{\mathrm{cell}} = \int_{\mathrm{cell}} \mathrm{d} V \rho$, where
$\rho$ is the local mass density. Consequently, the distribution of
particle masses is non-uniform and biased towards regions of high
density. This disparity is reduced by the logarithmic spacing of the
radial grid as zones at higher radii have in general both larger
volumes and lower densities.

The tracers record the evolution of density, temperature, radius,
electron fraction, neutrino luminosities, and energies. This allows us
to study the nucleosynthesis with the nuclear reaction network {\sc winnet }
\citep{wintelerphd, Winteler2012} that contains $6545$ nuclei up to
$Z=111$. The reaction rates are taken from the JINA Reaclib Database
V2.0 (\citealt{Cyburt2010}; accessed at 30/11/17) that is based on the
finite-range droplet mass model (\citealt{Moeller1995}). For nuclei
with $Z\ge 83$, we include neutron captures and neutron-induced fission from \cite{Panov2010} and $\beta$-delayed fission probabilities from \citet{Panov2005}. Neutrino reactions on nucleons are also included as in \citet{Froehlich2006}.

The nucleosynthesis calculations are performed for all tracers that are unbound at the end of the simulation. This set contains 6570, 7272, 17446, and 2218 particles for models 35OC-RO, 35OC-Rw, 35OC-Rs, and 35OC-RRw, respectively.  We start the network when the temperature of the tracers drops below $T=20$~GK. We assume nuclear statistical equilibrium (NSE) for 20~GK$>T>$7~GK and evolve only the weak reactions and the corresponding $Y_e$ variation\footnote{We observe deviations between the $Y_e$ from the hydrodynamical simulations and the one calculated in the network. This can become significant and depends on the initial temperature. Starting at a high temperature of $20\,\mathrm{GK}$ reduces these discrepancies.}. If the maximum temperature of a tracer is below 7~GK, we do not start from NSE but use the progenitor composition. For $T<7$~GK, the full network gives the detailed evolution of the abundances of each isotope. We run it until 1~Gyr, when most of the nuclei have decayed to stability. The tracers are extrapolated assuming an adiabatic expansion and density evolution as $\rho \propto t^{-3}$.


\section{Ejecta dynamics and nucleosynthesis}
\label{sec:results}

\begin{figure}
\begin{center}
  \includegraphics[width=0.95\linewidth]{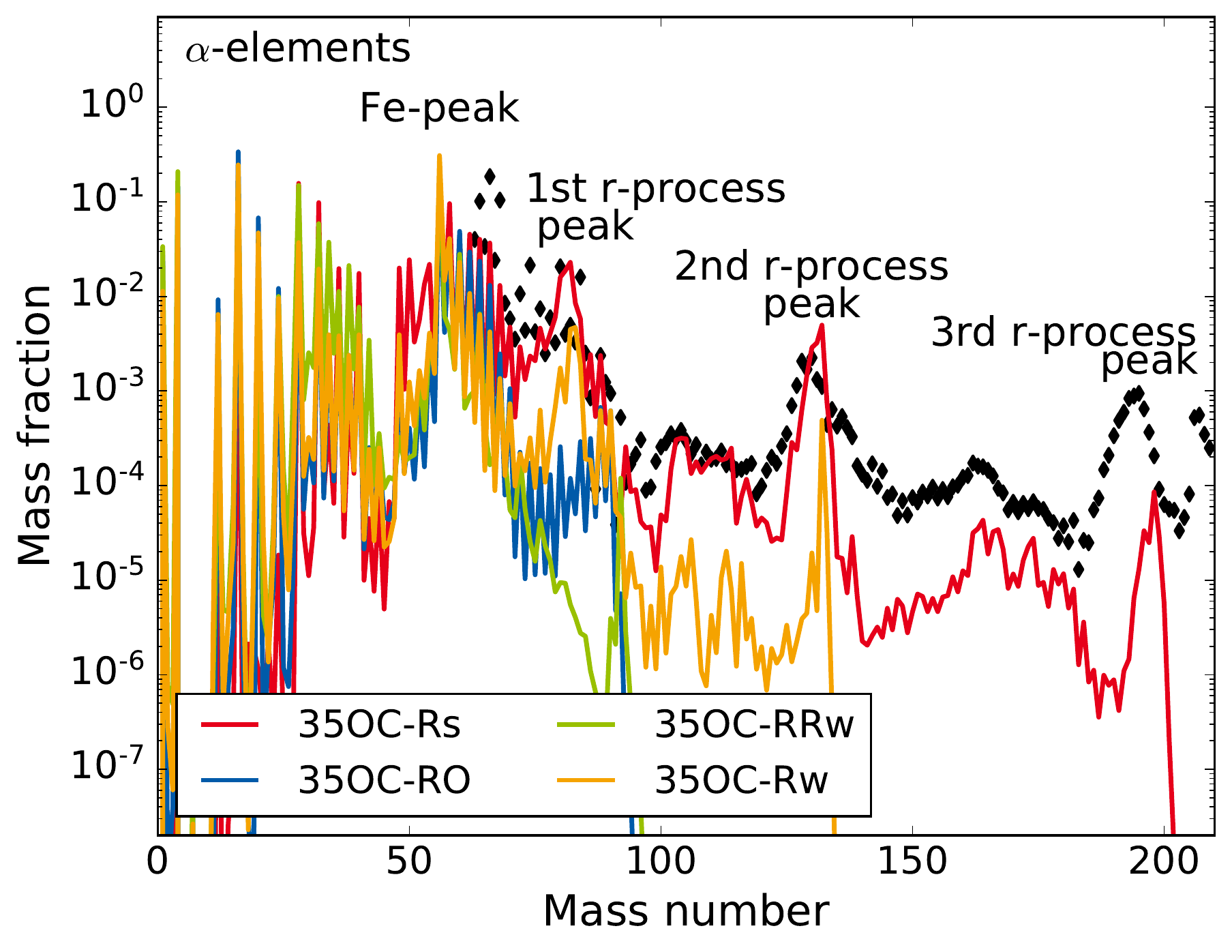}%
\end{center}
\caption{Integrated nucleosynthetic yields for the four models (see
  Sect.~\ref{sec:models} and Table~\ref{tab:models}) corresponding to different rotation velocities and magnetic fields. The black diamonds show the solar r-process residual \citep{Sneden2008}, normalized to mass number $\mathrm{A}=88$.}
\label{fig:integrated_pattern}
\end{figure}

The  ejecta composition taking into account all tracers is presented in Fig.~\ref{fig:integrated_pattern} for the models introduced in Sect.~\ref{sec:models}. The differences in the abundance patterns indicate that these models cover a wide range of nucleosynthesis conditions allowing to explore the impact of rotation, magnetic fields, and neutrinos.  The models 35OC-RO and 35OC-RRw are close to typical supernova explosion and produce also ``standard'' nucleosynthesis, namely elements up to the iron group and a bit of lighter heavy elements around $A\sim 90$ \cite[see e.g.,][]{Harris2017,Eichler2017,Wanajo2018,Ebinger2020}. The model with strong magnetic fields (35OC-Rs) synthesizes elements up to the third r-process peak \citep[see also][]{Nishimura2006,Winteler2012,Saruwatari2013,Nishimura2015,Nishimura2017,Halevi2018,Moesta2018}. The model 35OC-Rw is peculiar due to long-time evolution features that trigger the late ejection of neutron-rich material, see Sect.~\ref{sec:nuc_models_rot}.

\subsection{Nucleosynthesis patterns and hydrodynamical conditions}
\label{sec:classifying}

\begin{figure}
\begin{center}
\includegraphics[width=1.0\linewidth]{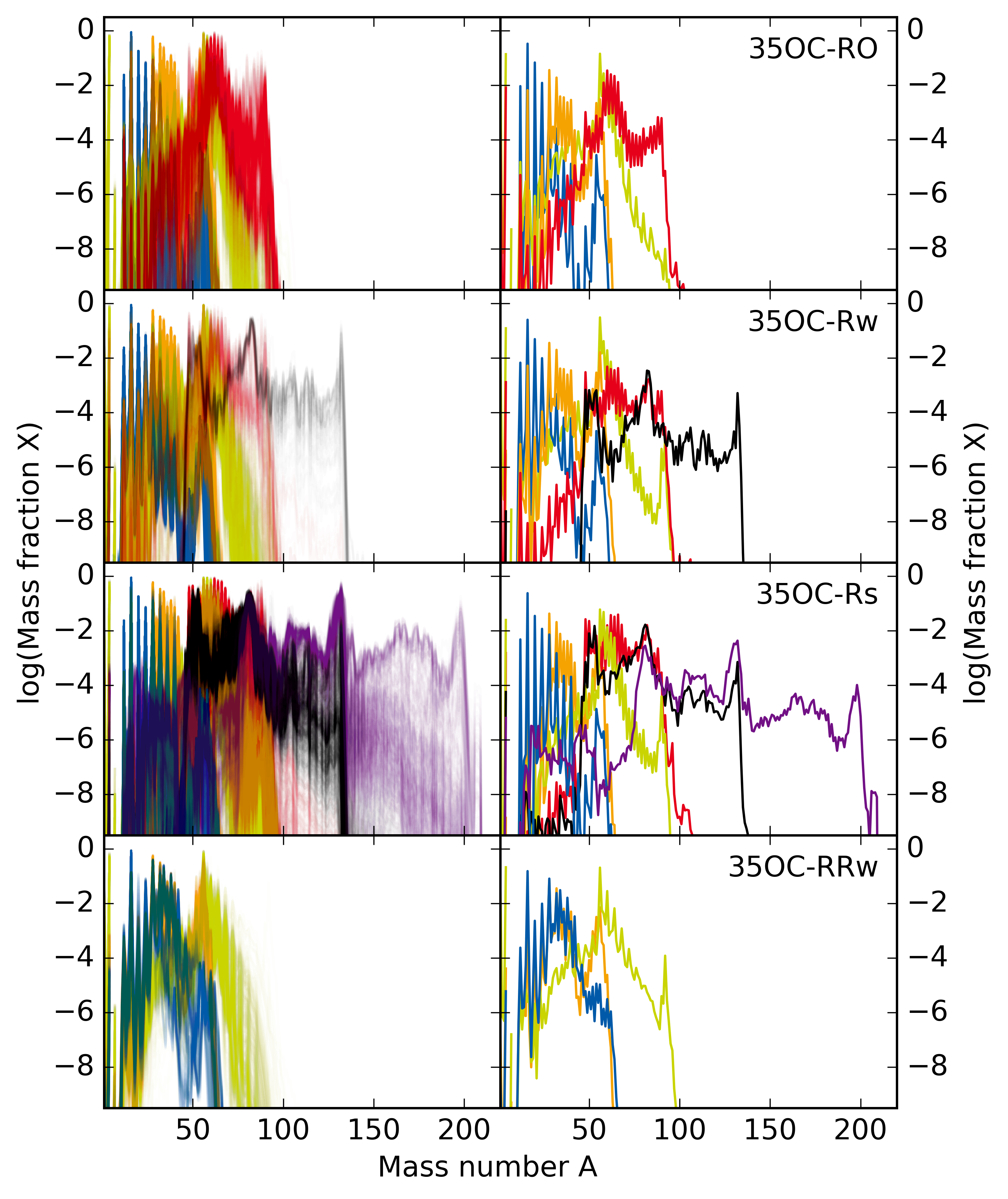}%
\end{center}
\caption{Composition of all  ejected tracer particles, indicated by one color for each group, binned by a k-means clustering algorithm. Left-hand panels: Final mass fractions after decay for each individual tracer. Right-hand panels: Mass-weighted integrated composition separated into the different groups.}
\label{fig:pattern_average_kmeans_maxtemp}
\end{figure}

\begin{figure}
\begin{center}
\includegraphics[width=1.0\linewidth]{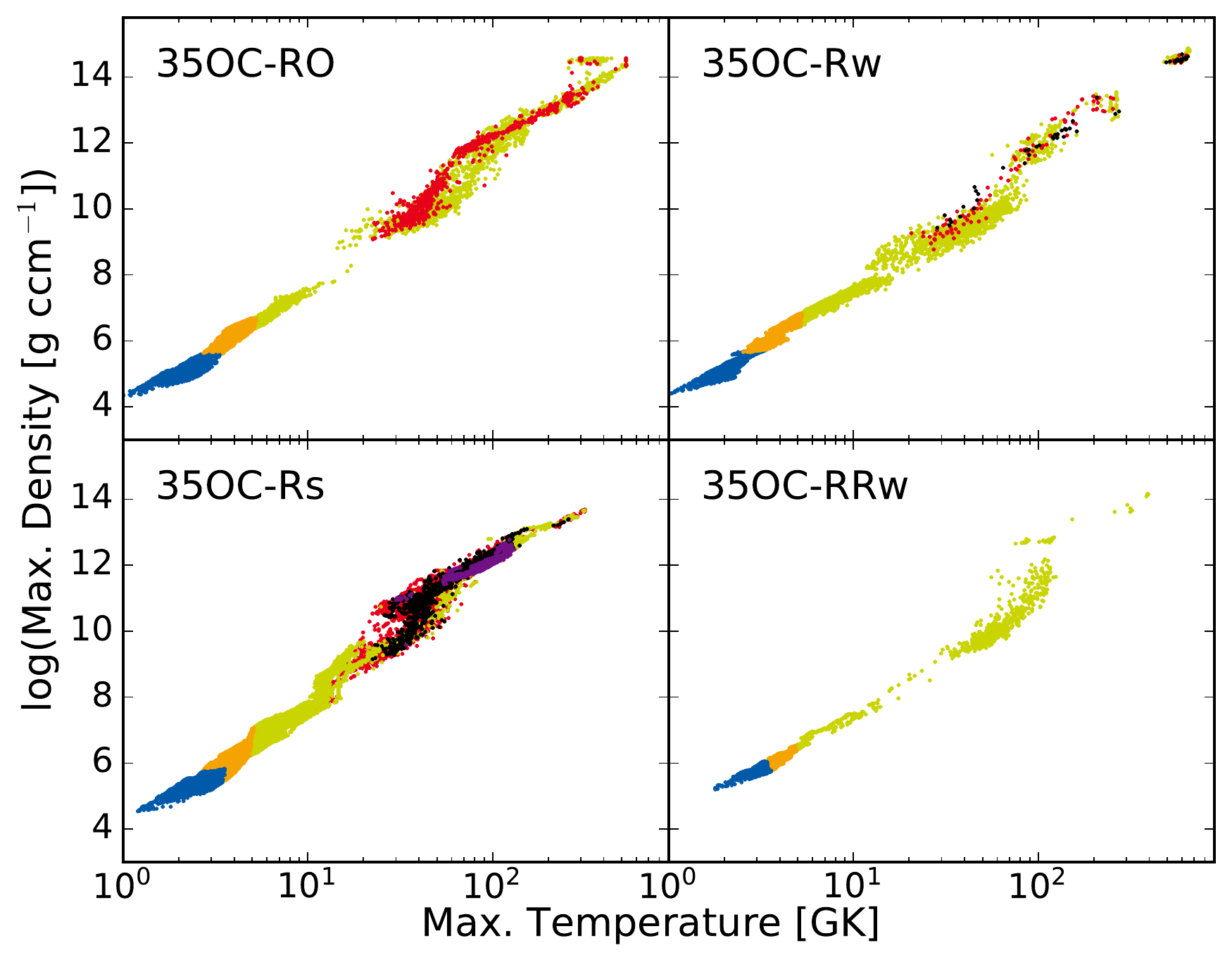}%
\end{center}
\caption{Maximum temperature and density of each tracer particle. The colours indicate nucleosynthetic groups as described in Sect.~\ref{sec:classifying}.} 
\label{fig:max_temp_dens}
\end{figure}

\begin{figure}
\begin{center}
\includegraphics[width=1.0\linewidth]{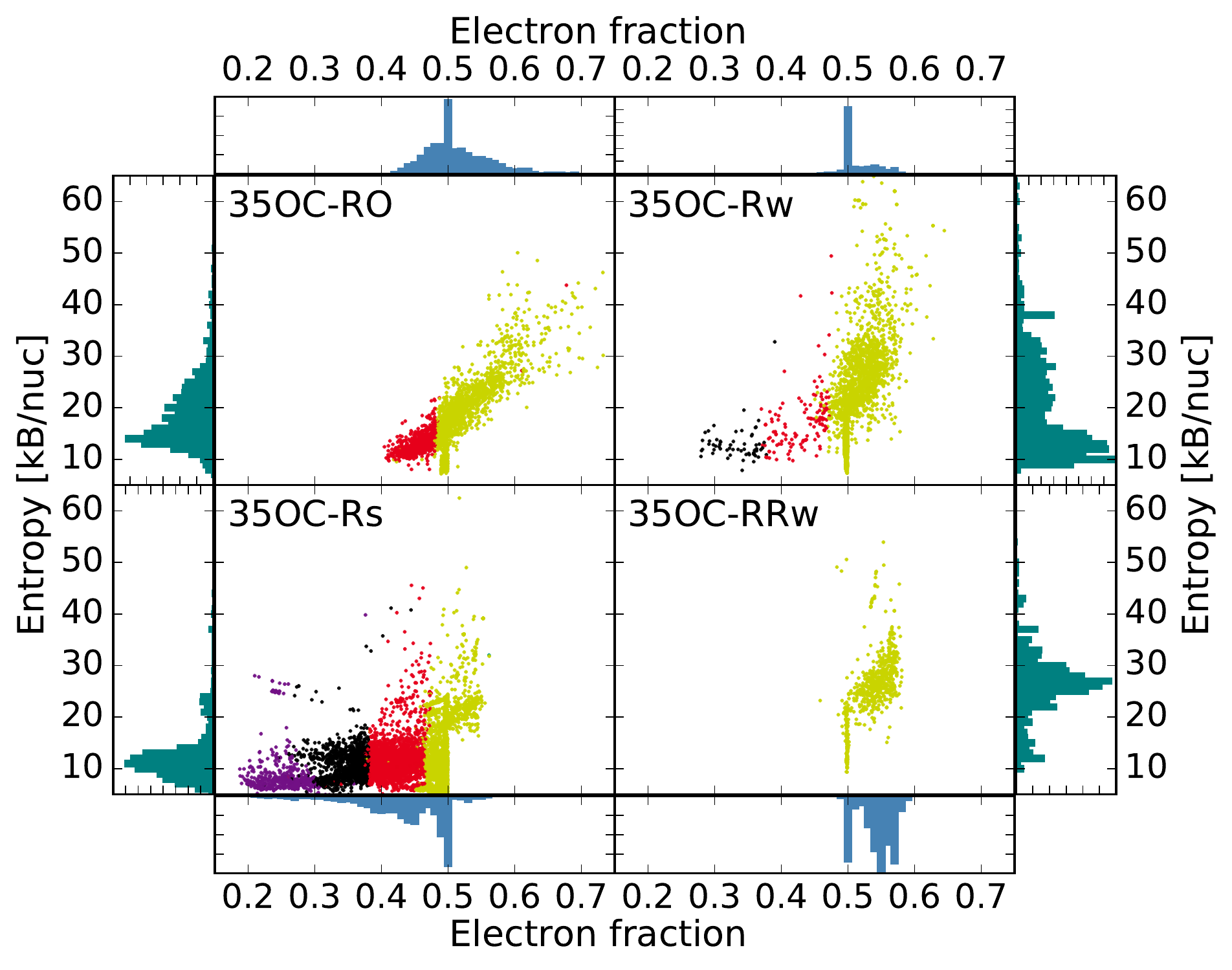}%
\end{center}
\caption{Electron fraction and entropy for each tracer at $T=5.8\;\mathrm{GK}$ together with their mass-weighted distributions at the outermost panels. Colours indicate the corresponding nucleosynthetic groups with the same color code as Fig.~\ref{fig:pattern_average_kmeans_maxtemp} and Fig.~\ref{fig:max_temp_dens}. Note that we only include tracer particles that reach a peak temperature of at least 5.8~GK here.} 
\label{fig:ye_entropy}
\end{figure}

\begin{figure}
\begin{center}
     \includegraphics[width=0.99\linewidth]{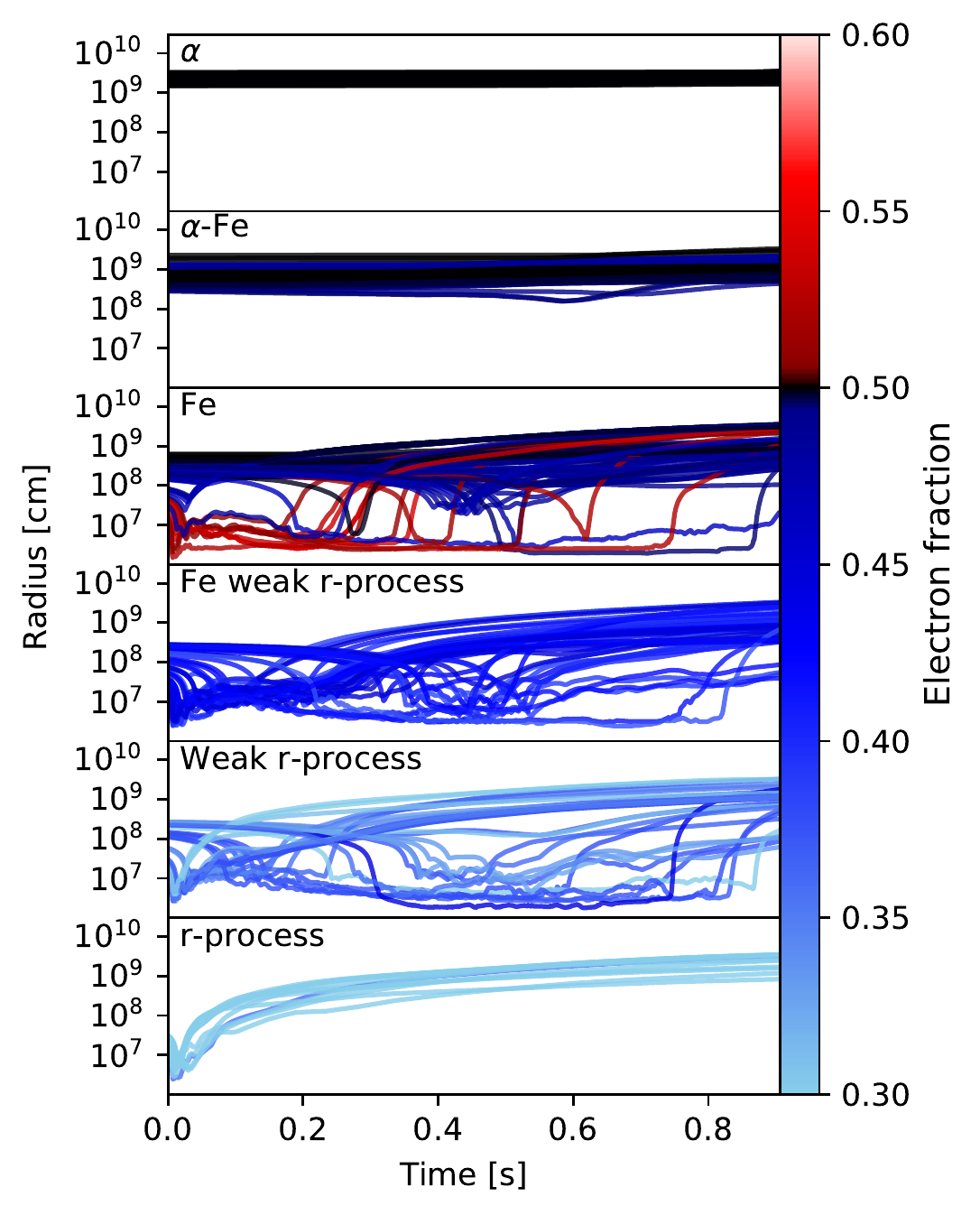}
\end{center}
\caption{Radial evolution of tracers from different groups of model 35OC-Rs. The colours indicate the $Y_e$ of the tracers at 5.8~GK or in the case of cooler trajectories the point of maximum temperature.}
\label{fig:radius}
\end{figure}

In order to understand the integrated abundances, we explore the
hydrodynamical conditions of individual tracer particles and the
corresponding nucleosynthesis contribution. The composition of every
tracer particle is shown in the left-hand panels of
Fig.~\ref{fig:pattern_average_kmeans_maxtemp}. Groups of tracer
particles with similar conditions lead also to similar abundance
patterns. We have separated these groups with the help of a k-mean
clustering algorithm \citep{Lloyd1982} for the abundances. The six
groups are indicated by different colours and the average composition
of each is shown in the right-hand panels of
Fig.~\ref{fig:pattern_average_kmeans_maxtemp}. Moreover,
Figure~\ref{fig:max_temp_dens} indicates the maximum density and
temperature of each tracer showing the strong link between the tracer
evolution and its nucleosynthesis. Every panel of this figure
corresponds to one of the four models (see Table~\ref{tab:models}) and
every dot to a tracer with the colours being the same as for the abundances
in Fig.~\ref{fig:pattern_average_kmeans_maxtemp}. The evolution
relevant for nucleosynthesis can be explained by the nucleosynthesis
parameters~\citep{Qian1996, Thompson2001} entropy and electron
fraction shown in Fig.~\ref{fig:ye_entropy}.  Although the
classification has been done based on abundance patterns, there is a
clear dependence of the groups on the electron fraction. The
histograms show the mass-weighted distributions of entropy and
 electron fraction. In all models, there is a peak around
$Y_e=0.5$. In addition to the entropy and electron fraction, the nucleosynthesis also depends on the expansion time-scale when the temperature drops down to around $T\approx 0.5$~MeV. Figure~\ref{fig:radius} shows the evolution of the radius for trajectories from different groups. There are three distinguished expansions: (1) trajectories that cross the shock and stays at large radii without approaching the neutron star, (2) trajectories that are promptly ejected from the outer layers of the PNS (bottom panel), and (3) trajectories that approach or even stay for some seconds close to the PNS and are ejected after being exposed to neutrinos.

In the following, we describe the six nucleosynthesis groups. The characteristics of a given group are the same for all models containing it, but only model 35OC-Rs contains all groups.

The \textbf{$\boldsymbol{\alpha}$ group} (blue in Fig.~\ref{fig:max_temp_dens}) contains tracers that are located
at large radii, which means that they encounter the shock at late
times, just before the end of the simulation (see
Fig.~\ref{fig:radius}). Therefore, they do not change much in the course of their
evolution and their maximum density and temperature do not exceed
$\rho = 10^6\;\mathrm{g \, cm^{-3}}$ and $T = 3\; \mathrm{GK}$,
respectively. Their electron fraction and final nuclear composition
resembles the original progenitor values, i.e. mainly
$\alpha$-elements.

The \textbf{$\boldsymbol{\alpha}$-Fe group} (orange) corresponds to moderately heated
progenitor material. The tracers of this group cross the shock at
earlier times when the latter is more energetic. Therefore, the peak
densities and temperatures are higher than for the $\alpha$ group
($\rho _\mathrm{peak} \approx 10^7\;\mathrm{g \;cm^{-3}}$ and
$T_\mathrm{peak} \approx$ 5~GK), leading to enhanced iron-group
abundances due to explosive burning, alongside a considerable amount
of $\alpha-$elements.

These two groups ($\alpha$ and $\alpha$-Fe) do not reach NSE
conditions, and they are characterized by their peak quantities. This
is visible by the clear and sharp separation of the groups in
Fig.~\ref{fig:max_temp_dens} compared to the other groups whose
maximum values overlap. Therefore, for the groups described below, the
maximum values are not determining the final abundances. Moreover,
within a same group, i.e., same abundance pattern, there are tracers
that seem to belong to different classes when viewed from a dynamic
perspective. Especially in the case of the Fe, Fe-weak-r-process, and
weak r-process groups, we can distinguish two subgroups of tracers 
defined by the time they start their outward propagation at high speeds (see Fig.~\ref{fig:radius}).

One subgroup consists of
tracers ejected shortly after they fall through the expanding shock
wave (in the following denoted the \emph{shock} subgroup) and another
one contains tracers ejected after passing many dynamical time-scales
in the vicinity of the PNS (\emph{inner} subgroup).

The \textbf{Fe group} (green) clearly contains the two subgroups and
the separation between both is visible in
Fig.~\ref{fig:max_temp_dens}, especially for models 35OC-RO and 35OC-Rw
for densities around $10^8\;\mathrm{g \;cm^{-3}}$. The shock subgroup
does not reach high values of the maximum temperature and density
because matter does not approach the neutron star. This results in
the electron fraction not changing much from the progenitor values and
staying around $Y_e\sim 0.5$ or slightly below. In contrast, the
tracers in the inner subgroup get close to the neutron star and are
ejected later (Fig.~\ref{fig:radius}). These tracers correspond to
the Fe-group points with higher maximum densities and temperatures in
Fig.~\ref{fig:max_temp_dens} and with $Y_e>0.5$ in
Fig.~\ref{fig:radius} (the black- and red-coloured lines).  In general, all tracers in both subgroups reach  peak temperatures high enough to photo-disintegrate the progenitor
composition and reach NSE. After NSE, their $Y_e$ distribution extends
from slightly neutron-rich to proton-rich conditions
($0.48 \le Y_e \le 0.6$; Fig.~\ref{fig:ye_entropy}). Under these
conditions, nuclear reactions favour the production of $^{56}$Ni that
later decays to $^{56}$Fe, as well as lighter heavy elements up to Zr
and Mo. This corresponds to typical nucleosynthesis found in
neutrino-driven explosions without rotation and magnetic
fields \citep[e.g.,][]{Harris2017,Eichler2017,Wanajo2018}. 
Notice that the proton-rich ejecta were not discussed in previous studies of MHD simulations because those did not include a
detailed neutrino transport to accurately account for this.

The \textbf{Fe-weak-r-process group} (red) is slightly more neutron
rich than the Fe-group with $0.38 \lesssim Y_e \lesssim 0.48$
(Fig.~\ref{fig:ye_entropy}). Most of the tracers come close to the
neutron star (inner group) and are ejected relatively fast, thus keeping
the neutron-richness of the outer layers of the PNS
star. Under such conditions the weak r-process produces lighter heavy
elements from Sr to Ag. Final abundances reach $A \sim 100$ and
are characterized by low abundances for alpha elements and high for
iron group elements

The \textbf{weak r-process group} (black) is dominated by the inner
subgroup of tracers with maximum temperatures of
\mbox{$T_\mathrm{max}>20\; \mathrm{GK}$} and densities of
\mbox{$\rho _\mathrm{max}=10^9\;\mathrm{g\; cm^{-3}}$}. These values
are similar to the ones of  the Fe-weak-r-process group but $Y_e$ is lower
(Fig.~\ref{fig:ye_entropy}) due to a faster expansion and thus shorter
exposure to neutrinos in the expansion phase. These conditions favour
the production of elements up to the second r-process peak around
$A\sim 130$. This group is not a robust feature of all models.
Instead, its presence depends on special conditions that are met only
in two models, 35OC-Rs and, to a lesser degree, 35OC-Rw.  In the
former model, weak r-process tracers are ejected at all times, whereas
in the latter only a unique transformation of the PNS causes them to
appear at very late times (see Sect.~\ref{sec:nuc_models_rot}).

The matter in the \textbf{r-process group} (purple) promptly and
quickly accelerates from the neutron star surface and along the jets
as indicated by the radius evolution shown in
Fig.~\ref{fig:radius}. This fast expansion prevents that the neutrinos
transform neutrons into protons resulting in low electron fractions
(Fig.~\ref{fig:ye_entropy}). This group contains nuclei heavier than A
$\geq$ 130 and reaches the third r-process peak (A $\sim$ 195). The
neutron-rich material of this group that is ejected very early along the jets shifts to the sides of the jet at later times. The late
configuration consists of proton-rich jets surrounded by neutron-rich
clumps where the r-process occurs.

\subsection{Impact of rotation and the weak r-process}
\label{sec:nuc_models_rot}

The effect of rotation can be investigated by comparing the two models
with similar weak magnetic fields: 35OC-Rw and 35OC-RRw. Both models
produce abundances for alpha elements and up to the iron
group\footnote{Note that the outer layers of the progenitor are not
  included here and they contribute to the alpha elements, see e.g.,
  \citet{Eichler2017}.}.

\begin{figure}
\begin{center}
\includegraphics[width=1.0\linewidth]{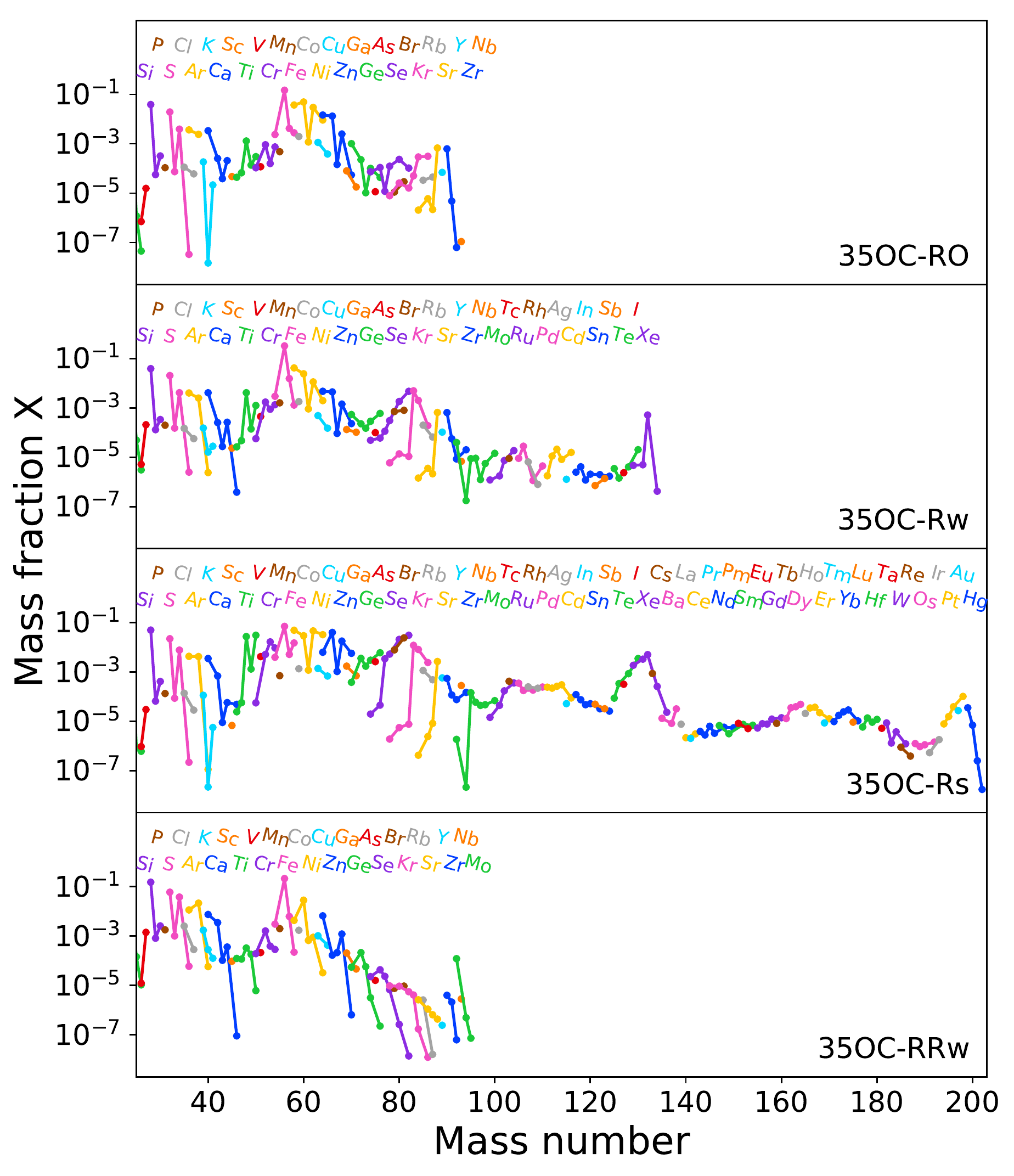}%
\end{center}
\caption{Mass fractions of individual isotopes for every model. Isotopes of a same
  elements are indicated by a given color and connected by a line. The
  element names are given at the top of each panel. Nuclei with mass
  fractions $\leq 10^{-8}$ are not included.}
\label{fig:isotopes}
\end{figure}

Model 35OC-RRw with strong rotation and weak magnetic field is
characterized by only proton-rich ejecta in addition to
the $\alpha$ and $\alpha$-Fe groups. Rotation reduces the accretion and thus the
accretion luminosity, and this makes the explosion slower and matter
stays  exposed to neutrinos for a longer time. The result is that the
ejecta are proton rich as shown in Fig.~\ref{fig:ye_entropy}. Here, we
find typical nucleosynthesis produced by the $\nu$p-process
when the matter flow runs on the proton-rich side of stability
\citep{Froehlich2006, Pruet2006, Wanajo2006}. In addition, for conditions with $Y_e \sim 0.5$ or slightly proton- or neutron-rich,
the flow goes along stability. The proton-rich conditions produce characteristic isotopic abundances including  p-nuclei as shown in the bottom panel of Fig.~\ref{fig:isotopes}, see \cite{Bliss2018, Eichler2017}, and \cite{Wanajo2018} for more details about the nucleosynthesis in proton-rich supernova ejecta.

\begin{figure}
\begin{center}
\includegraphics[width=1\linewidth]{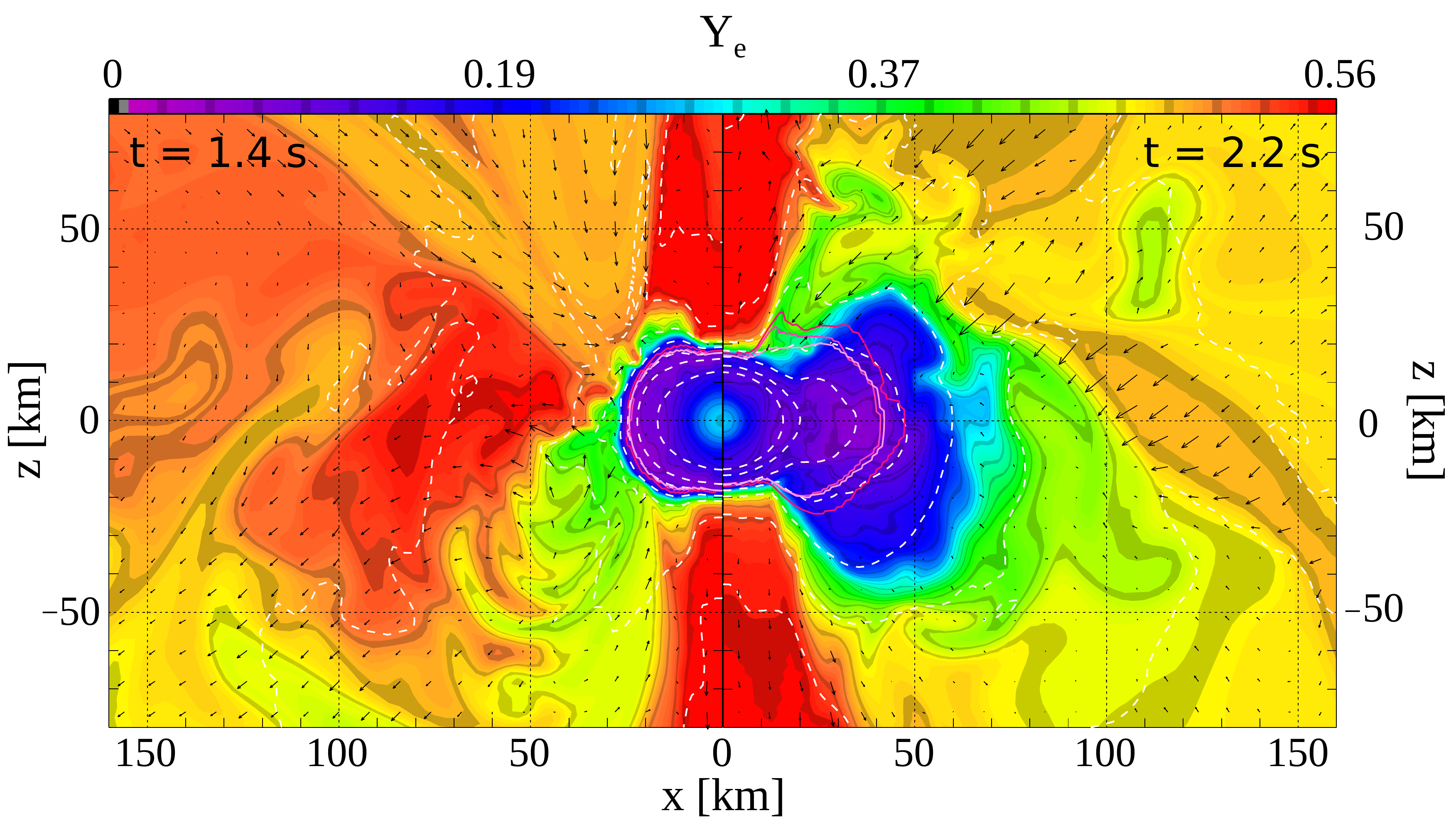}%
\end{center}
\caption{Electron fraction of model 35OC-Rw in a region
  around the PNS at $t \sim1.4 \; \mathrm{s} $ (left-hand panel) and 2.2~s (right-hand panel). Contours of constant density ($10^{14},10^{13},..\;\mathrm{g cm}^{-3}$) are indicated with the white, dashed lines. The pink contours correspond to the neutrinospheres.}
\label{fig:ye_rw_explanation}
\end{figure}

In the model with slower rotation (35OC-Rw), most of the matter is
ejected with $Y_e \sim 0.5$ and a small amount is slightly neutron
rich and the weak r-process produces the lighter heavy elements up to
around Ag \citep[see e.g.,][]{Bliss2017}. In addition, there is a late matter
ejection ($t \gtrsim 2 \, \mathrm{s}$) with $Y_e \sim 0.3$. The sudden appearance of such a population of tracers is the
consequence of a relatively abrupt change in the PNS structure that
had occurred slightly earlier. Up to $t \sim 1.4 \; \mathrm{s}$, the
PNS is almost spherical with a decreasing radius and an aspect ratio
close to unity despite having a very high rotational
energy. Eventually, however, its magnetic field grows sufficiently to
redistribute angular momentum to the outer layers. The excess
centrifugal support causes these layers to expand and leads to a
growth of the ratio between equatorial and polar radius beyond a value
of two (Fig.~\ref{fig:ye_rw_explanation}). This expansion affects
matter of very low $Y_e$ (marked by the blue colours in the figure), some
of which even ends up outside the neutrinospheres. The turbulent
fluid flows in this region stochastically advect parcels of this very
neutron-rich matter into the polar outflows. These fluid elements will
be ejected at very high speeds and $Y_e$ stays low
(Fig.~\ref{fig:ye_entropy}).  We note that no similar transition from
a spherical to an oblate PNS takes place in model 35OC-RO. There, the
magnetic field is strong enough to cause a high aspect ratio already
early on. Although we find neutron-rich matter outside the
neutrinospheres also in this case, the amount is less and the
structure of the PNS makes it less likely for this matter to enter the
outflow, thus suppressing the weak r-process group.
  
\subsection{Impact of the magnetic field and the r-process}
\label{sec:nuc_models_B}
Models 35OC-Rw, 35OC-RO, and 35OC-Rs show the impact of increasing
magnetic field strengths on the abundances
(Fig.~\ref{fig:integrated_pattern}). When increasing the magnetic
field from model 35OC-Rw to model 35OC-RO, then elements around the
second r-process peak are not produced anymore. This is related to the
late evolution of model 35OC-Rw, discussed above. We note, however,
that this non-monotonicity, caused by the presence or the absence of late
neutron-rich fluid elements, only affects a small fraction of the
ejecta. When these fluid elements are ignored, the distribution of the
ejecta across $Y_e$ behaves monotonically with initial magnetic field strength
(Fig.~\ref{fig:ye_entropy}).
  
Explosions with strong magnetic fields, like 35OC-Rs, have been
suggested as a potential r-process site \citep[e.g.,
][]{Meier1976,Meyer1994, Nishimura2006, Winteler2012,Nishimura2015,Nishimura2017,Moesta2018}. The magnetic field produces a jet-like
explosion and prompt ejection of neutron-rich material
(Fig.~\ref{fig:radius}). The formation and stability of the jets
strongly depends on the magnetic field and is still under discussion
\citep{Moesta2014,Kuroda2020,Obergaulinger_Aloy_2020__mnras_PaperIII}.  For strong magnetic fields, some neutron-rich
matter is rapidly ejected by the magnetic pressure without major
neutrino interactions with neutrons. Depending on whether the magnetic
field or neutrinos dominate as ejection mechanism, we find a strong
r-process or a weak r-process, respectively. This behaviour has been
investigated in detail by, e.g. \citet{Nishimura2017} and \citet{Moesta2018}
when they artificially varied the neutrino luminosity. Our neutrino
treatment is self-consistent and the neutrino luminosity is not a free
parameter.

In model 35OC-Rs, matter close to the neutron star and thus at small
radii (Fig.~\ref{fig:radius}) reaches high densities and temperatures
(Fig.~\ref{fig:pattern_average_kmeans_maxtemp}) and is promptly
ejected. Due to the fast
expansion, neutrinos are unable to convert too many neutrons into
protons and the electron fraction stays low, $Y_e=0.2-0.3$
(Fig.~\ref{fig:ye_entropy}). Such conditions (i.e., fast expansion and
low $Y_e$) allow for the r-process to produce elements up to the third
r-process peak (Fig.~\ref{fig:pattern_average_kmeans_maxtemp}).

In addition to the r-process, model 35OC-Rs ejects matter with different conditions leading to a large range of nucleosynthesis
processes and products (Figs.~\ref{fig:pattern_average_kmeans_maxtemp}
and~\ref{fig:ye_entropy}). The jets are very
proton-rich and contribute  mainly to iron-group
elements but also to heavier ones by the $\nu$p-process \cite{Pruet2006,Froehlich2006}, and \cite{Wanajo2018}. Moreover, there is neutron-rich matter that is continuously
ejected around the jets and produces elements up to the second
r-process peak by a weak r-process. Remarkably, this neutron-rich matter comes from the PNS outer layers. It is extracted from there due to the mechanical action of the coherent, large-scale magnetic field in this model. As thoroughly discussed in \cite{Aloy_Obergaulinger_PaperII_2020arXiv200803779A}, this process is so strong that yields to a decrease in the PNS mass for $t\gtrsim 0.5\,$s post-bounce.

\section{Observables}
\label{sec:obs}

Here, we present a comparison of the nucleosynthesis produced in our models
with observations, which focuses on three aspects: (1) the iron group  element production that can be observed in supernova light
curves, (2) heavy radioactive isotopes that may be visible by
gamma- or X-rays, and (3) r-process elements compared to stellar elemental
abundances. Table~\ref{tab:nuc_summary} provides the ejected mass of representative isotopes and elements for every model to get an overview. A full list of the synthesized isotopes is given in Appendix~\ref{app:datatables} in Tables~\ref{tab:data_ro}, \ref{tab:data_rw}, \ref{tab:data_rrw}, and \ref{tab:data_rs}.

\begin{table}
 \begin{center}
  \caption{Explosion energy and yields of selected isotopes and elements for  different models.}
  \label{tab:nuc_summary}
  \begin{tabular}{rcccc}
    \hline
    &35OC-RO & 35OC-Rw & 35OC-Rs &35OC-RRw\\
        \hline
        $E_{\mathrm{exp}} [B]$ & 1.78 & 2.8 & 4.16 & 0.21 \\
      $^{26}$Al   & 2.26 (-7)      & 1.94 (-6)    & 3.62 (-7)        & 4.33 (-7)  \\
      $^{44}$Ti   & 6.60 (-5)      & 1.34 (-4)    & 2.06 (-5)        & 1.16 (-5)  \\
      $^{60}$Fe   &  4.94 (-4)     & 1.55 (-4)    & 3.62 (-3)        & 1.69 (-7)  \\
    $^{56}$Ni     & 4.73 (-2)      & 1.21 (-1)    & 2.54 (-2)        & 7.32 (-3)  \\
   $^{129}$I      & -              & 1.75 (-6)    & 6.93 (-4)        & -          \\
   $^{137}$Cs     & -              & -            & 3.18 (-6)        & -          \\
   $^{247}$Cm     & -              & -            & 2.30 (-12)       & -          \\  
    Mn            & 1.53 (-4)      & 6.23 (-4)    & 2.74 (-4)        & 6.87 (-4)  \\
    Zn            & 9.77 (-3)      & 4.23 (-3)    & 2.74 (-2)        & 2.81 (-3)  \\
    Sr            & 2.20 (-4)      & 2.56 (-4)    & 1.03 (-3)        & 1.65 (-6)  \\
    Y             & 2.22 (-5)      & 4.05 (-5)    & 2.23 (-4)        & 8.42 (-8)  \\
    Zr            & 2.01 (-4)      & 2.84 (-4)    & 3.45 (-4)        & 1.29 (-7)  \\
    Ba            & -              & 2.84 (-10)   & 2.07 (-5)        & -          \\
    Pr            & -              & -            & 7.94 (-7)        & -          \\
    Nd            & -              & -            & 1.07 (-5)        & -          \\
    Eu            & -              & -            & 5.19 (-6)        & -          \\
    Dy            & -              & -            & 5.29 (-5)        & -          \\
    Pt            & -              & -            & 6.39 (-5)        & -          \\
    Au            & -              & -            & 1.06 (-5)        & -          \\
    
  \hline
  \multicolumn{5}{p{\dimexpr\linewidth-2\tabcolsep-2\arrayrulewidth}}{Note: Yields in M$_{\odot}$ using the notation $A(B)$ for $A\times 10^B$. Radioactive isotope yields are given as maximum synthesized value. Note that $^{26}$Al and $^{60}$Fe are also synthesized during stellar evolution \citep[e.g.,][]{Limongi2006} and the progenitor contribution is not included here.}
  \end{tabular}
\end{center}
\end{table}

\subsection{Synthesis of Ni and Co.}
\label{sec:Ni}

The four models analysed give a good and reliable overview about the
iron group elements produced in the first seconds of neutrino-driven
and magneto-rotational-driven supernovae. Our calculations are the
first ones based on detailed neutrino transport and long-time
MHD simulations in 2D. We only investigate the nucleosynthesis of
the innermost ejecta without including the outer layers of the star. Even
if the simulations are followed for up to several seconds after the
explosion, there is still matter that will be ejected later and that is hot
enough to further contribute to the production of iron group
nuclei. Therefore, the results discussed in the following show trends
and lower limits for the mass ejected of given isotopes. The model
35OC-Rs provides a unique opportunity to study the early
nucleosynthesis of a MR-SN and compare to observed hypernovae and long
GRB-SNe. In addition, models 35OC-RO and 35OC-RRw are valuable examples
for standard neutrino-driven supernova nucleosynthesis including rotation and magnetic fields in the computational set-up.

The values that we find for $^{56}$Ni range from
$7.3\times 10^{-3}M_\odot$ for the fast-rotating and weakly exploding
model (35OC-RRw) to $1.2\times 10^{-1}M_\odot$ for model 35OC-Rw. The
amount of $^{56}$Ni and explosion energy correlates for the three
models without strong magnetic field (see
Table~\ref{tab:nuc_summary}). Model 35OC-Rs, with strong magnetic field, has
significantly larger explosion energy while the ejected $^{56}$Ni mass
is still low, although this may increase as nucleosynthesis is still
going on at the end of the simulation. In any case, it may be
unlikely that the amount of Ni still increases from
$2.5\times 10^{-2}M_\odot$ to larger than $\sim 0.3\,\Msol$ as predicted by
hypernova observations (see \citealt{Iwamoto_1998Natur.395..672_SN1998bw,Nakamura_2001ApJ...550..991N_SN1998bw, Mazzali_2003ApJ...599L..95_SN2003dh, Nomoto2006} and references therein). The explosion energy of this
model is also below the typical hypernova energy ($> 10$~B) but it may
still increase enough to become a low-energy hypernova. In general, our
 models produce explosion energies and $^{56}$Ni masses that are
similar to values from observations (e.g., for SN1987A
$M(^{56}\mathrm{Ni})\approx 7\times10^{-2}M_\odot$ from
\citealt{Seitenzahl2014}). Another radioactive isotope observed in
supernovae remnants is $^{44}$Ti with for example
$M(^{44}\mathrm{Ti})\approx 5.5\times10^{-5}M_\odot$ for SN1987A
\citep{Seitenzahl2014} and
$M(^{44}\mathrm{Ti})\approx 1.3\times10^{-4}M_\odot$ for Cas~A
\citep{WangLi2016}. Our yields are around these estimates, although
the exact observed value may be still uncertain (see
\citealt{INTEGRAL_Ti2012}, \citealt{Seitenzahl2014}, \citealt{Weinberger2020}) or the individual values of the compared stars may be exceptional \citep[which may be the case in SN1987A;][]{Podsiadlowski_1992PASP..104..717}.

\subsection{Radioactive isotopes}
\label{sec:rad_iso}

The decay of radioactive isotopes produced during the supernova
explosion can be observed by their gamma- and X-rays. This is a
direct observation of supernovae and stellar nucleosynthesis as it has
been done for $^{44}$Ti,  $^{26}$Al, $^{60}$Fe, and more \citep[see][for reviews]{Diehl1998,Diehl2007}. Heavy radioactive isotopes have not been observed yet
although this may become possible as suggested by previous studies
\citep{Qian1998,Ripley2014,Korobkin2020}.

Here, we briefly show the production of some radioactive isotopes in our models and their potential detection. The flux for a
given ejecta composition is \citep{Qian1998}
\begin{equation}
F_\gamma = \frac{N_\mathrm{A}}{4\pi d^2}\frac{M_x}{A} \frac{I_\gamma}{\bar{\tau}},
\end{equation}
where $N_\mathrm{A}$ is the Avogadro's number, $d$ the distance to the
remnant, $M_x$ the mass of nucleus $x$, $I_\gamma$ the
number of photons per energy emitted, and
$\bar{\tau}= T_{1/2}/\ln(2)$ the lifetime of the investigated
nucleus. For $I_\gamma$ we take values obtained from the Lund/LBNL
Nuclear Data Search\footnote{http://nucleardata.nuclear.lu.se/toi/, the database was compiled by S.Y.F. Chu, L.P. Ekström and R.B. Firestone.},
whereas we use lifetimes from Nuclear Wallet Cards \citep{Tuli2011}. Notice that our calculation is only an estimate, the obtained spectrum contains only emission lines, no continuum emission (e.g., by bound-free or free-free interactions), and no absorption. Furthermore, we did not take any line-broadening effect into account and when observed with a real telescope, the emission lines will have a distribution with finite width and therefore a lower maximum flux.

We investigate model 35OC-Rs and take only one representative trajectory of each nucleosynthetic group weighted by the mass of the corresponding group. Several studies and observations are available
for iron group (e.g., $^{44}$Ti) and lighter nuclei ($^{26}$Al, $^{60}$Fe). Here we find also those light isotopes (see Table~\ref{tab:nuc_summary}) and go beyond by looking at the emission from heavy r-process nuclei. 

\begin{figure}
\begin{center}
\includegraphics[width=1.0\linewidth]{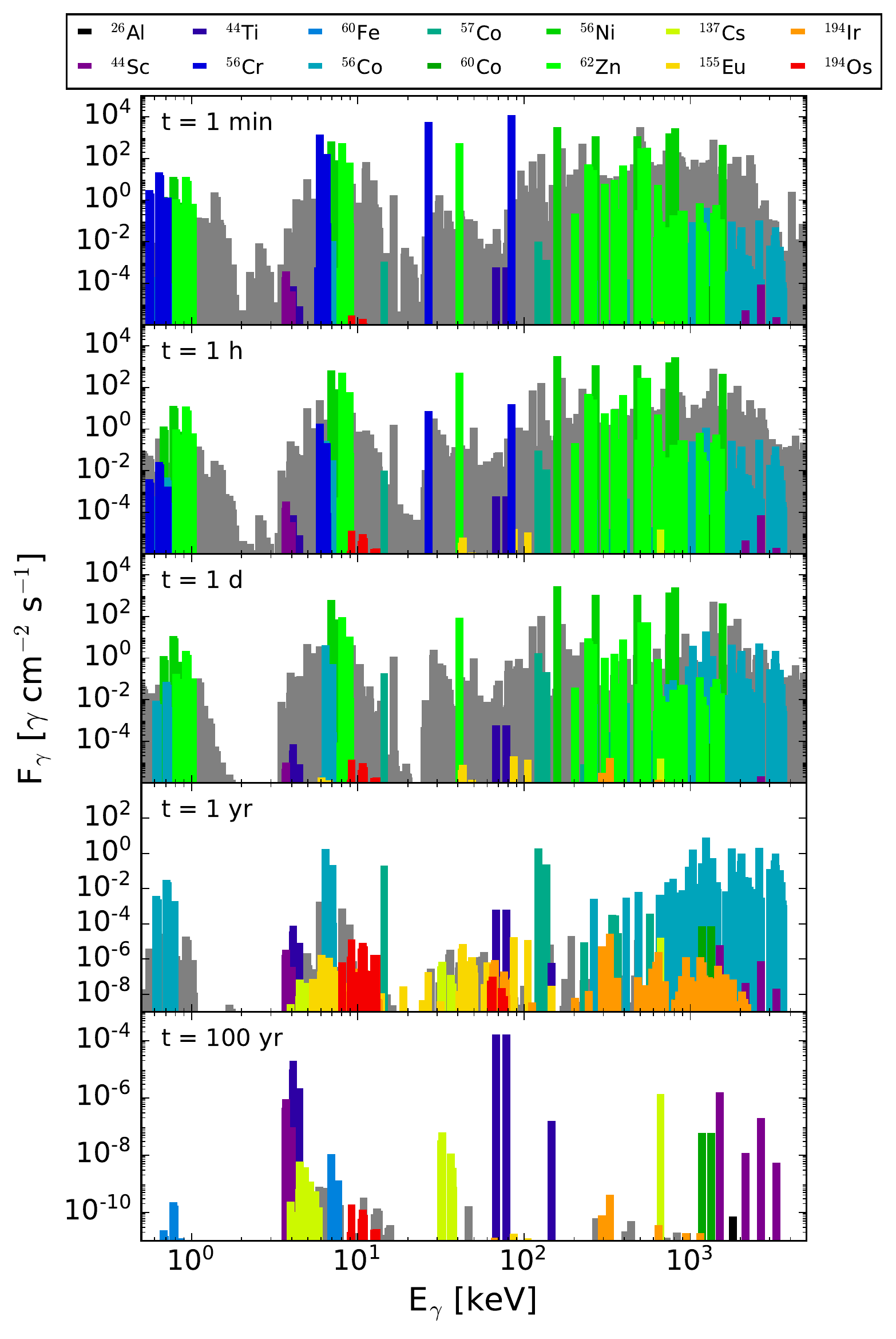}
\end{center}
\caption{Flux at 1\, kpc for different emission lines of model 35OC-Rs. Each panel corresponds to different times.} 
\label{fig:fluxshorttime}
\end{figure}

The time when to detect the gamma- and X-rays is critical. At early times after the event, the total flux will be higher. Due to the still high velocities of the ejecta in addition to large and slowly decreasing opacities, the detection will, however, be challenging. Nevertheless, there are studies that identified peculiar features of individual emission lines in the afterglow of GRBs \citep{Margutti2008,Campana2016}. Remarkebly, \citet{Margutti2008} found an emission line around $E_\gamma\sim 7.85$\, keV. This coincides with the emission lines of Ni and Cr radioactive isotopes at around these energies at early times (Fig.~\ref{fig:fluxshorttime}). Similarly, the spectrum shows a feature around $E_\gamma\sim 0.5$\, keV, which again agrees with emission lines of iron group nuclei in the spectrum. Heavier r-process elements lead to lower fluxes than iron group nuclei that dominate the spectrum. Therefore, it does not seem likely to make a direct detection of the r-process in the spectrum at early times. 

Another possibility for a detection can be achieved by looking at later times at SN remnants \citep{Qian1998,Ripley2014}, where short-lived radioactive isotopes have already decayed. In the remnants of SN1987A and Cassiopeia A,   $^{44}$Ti emission lines were detected around $E_\gamma\sim 68$ and $\sim 78$\, keV \citep{INTEGRAL_Ti2012,Grefenstette2014}. These emission lines are also visible in the lower panels of Fig.~\ref{fig:fluxshorttime}. When focusing on r-process elements, there exists emission lines of $^{137}\mathrm{Cs}$, $^{155}\mathrm{Eu}$, $^{194}\mathrm{Ir}$, and $^{194} \mathrm{Os}$ (lower panels of Fig.~\ref{fig:fluxshorttime}). The feature of $^{137}\mathrm{Cs}$ maintains a relatively high flux even after an extreme time of $100$\, yr.

\begin{figure}
\begin{center}
\includegraphics[width=1.0\linewidth]{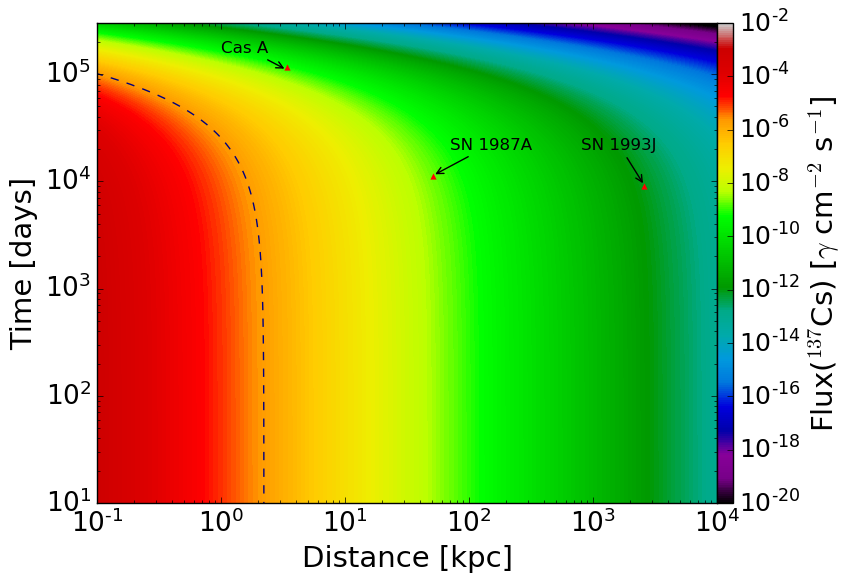}
\end{center}
\caption{Flux of the $^{137}$Cs gamma-ray line at $E_\gamma=661.7\, \mathrm{k eV}$. Shown is time versus distance of the source. The dashed line indicates a constant flux of $2\times 10^{-6} \, \mathrm{\gamma \, cm^{-2}\, s^{-1}}$, which is a flux that could be detected by gamma-ray telescopes as AMEGO \citep{Rando2017} or e-ASTROGAM \citep{DeAngelis2017} at $E_\gamma \approx 1\, \mathrm{M eV}$. Data for Cas A were taken from \citet{Ferrand2012,Green2009}, for SN1987A from \citet{Panagia2003}, and for SN 1993J from \citet{Freedman1994}.} 
\label{fig:Cs_obs}
\end{figure}

Fig.~\ref{fig:Cs_obs} shows the time and distance for known supernova remnants together with the expected
flux from the decay of  $^{137}$Cs from a MR-SN at given distance and
age. As indicated by the dashed line, only remnants within $\sim 3$~kpc
may provide a significant signal. Whether this features can be observed with upcoming detectors is beyond the scope of our work.

\subsection{r-process and UMP stars}
\label{sec:sneden_vs_honda}

\begin{figure}
\begin{center}
\includegraphics[width=1.0\linewidth ]{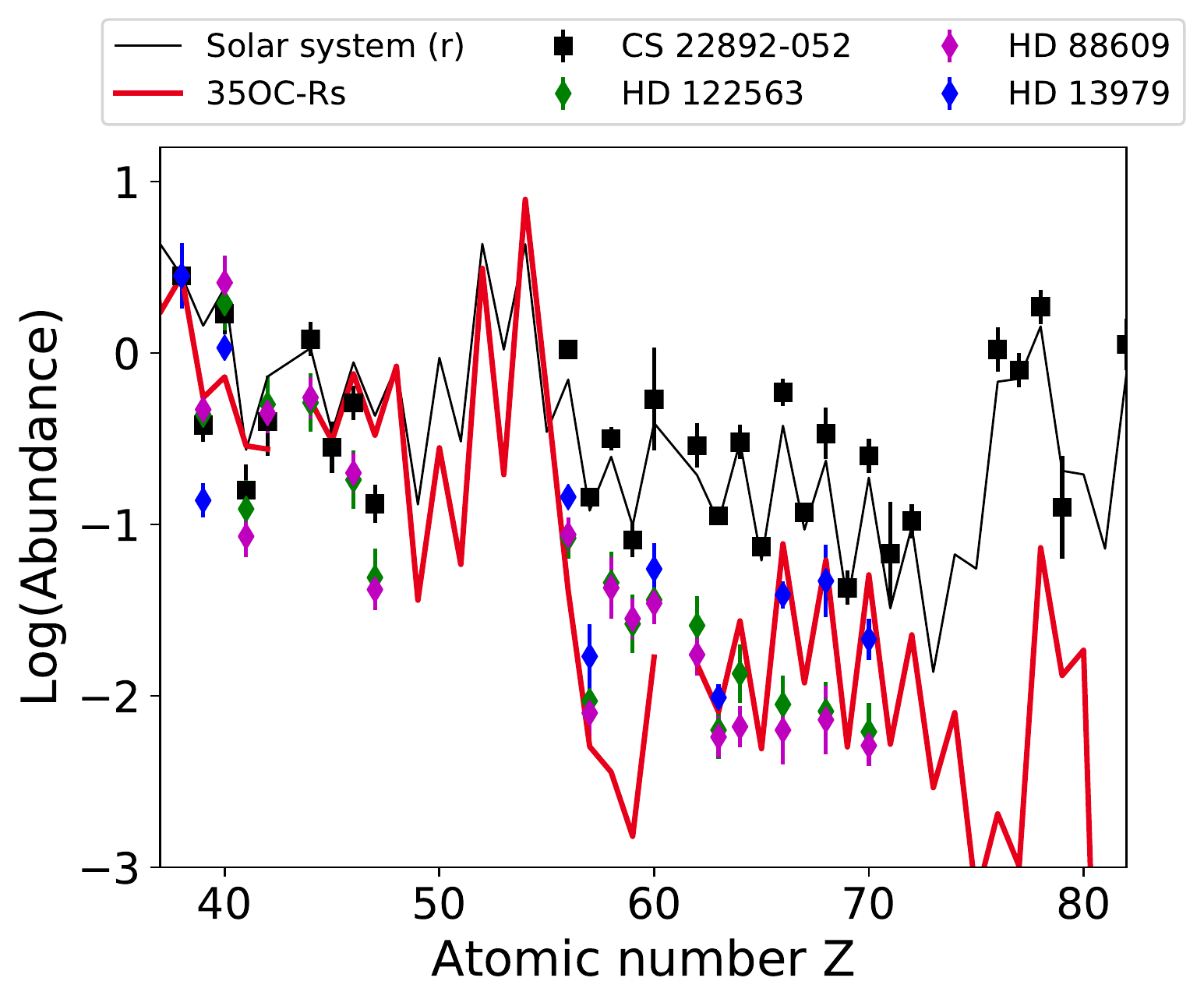}%
\end{center}
\caption{Comparison of the final abundances to the r-process enriched
  star CS 22892-052 \citep{Sneden1996}, HD 122563
  \citep{Honda2006}, HD 88609 \citep{Honda2007}, and HD 13979 \citep{Roederer2014}. We have normalized all abundances to  strontium ($\mathrm{Z}=38$) of model 35OC-Rs.}
\label{fig:sneden_honda_comp}
\end{figure}

Ultra-metal-poor (UMP) stars belong to the oldest stars and thus they
provide a unique possibility to study the nucleosynthesis from MR-SNe produced from sub-solar metallicity stellar progenitors and validate our models against observations (see also
\citealt{Nishimura2017}). The elemental abundances observed in the atmosphere of such old stars come from few previous nucleosynthesis events. We compare observed abundances to our models in Fig.~\ref{fig:sneden_honda_comp}. Usually there are two types of abundance patterns with high and low enrichment of the elements between second and third r-process peaks (see e.g.,~\citealt{Qian2001,Qian2007,Qian2008, Hansen2014}). Most of the stars with high
enrichment of heavy r-process elements present a robust pattern for
those, meaning that the relative abundances among elements are very
similar among different stars and also follow the solar r-process  \cite[see
e.g.,][]{Sneden2008}. These stars are sometimes called ``Sneden-like'' stars. For the lighter heavy elements below the second peak, even in Sneden-like stars, there is more variability or less robustness in the
patterns.  In addition, there are many different patterns for stars with low enrichment of heavy r-process elements \citep{McWilliam1998,Aoki2005,Honda2006, Roederer2010}. These are called ``Honda-like'' stars.

In Figure~\ref{fig:sneden_honda_comp}, we show how our models can
explain different observational features. Models 35OC-RO and 35OC-RRw
contribute to the lighter heavy elements as expected from standard
supernova nucleosynthesis \citep{Harris2017, Bliss2018,Eichler2017, Wanajo2018}. This supernova
contribution to the lighter heavy elements can give an explanation to the variability of the abundance patterns for those elements \citep{Qian2001,Sneden2008, Hansen2014}. Model 35OC-RW, with its peculiar late evolution,
reaches the second r-process peak. This model can provide some hints
to the weak r-process production of elements beyond Sr, Y, Zr but
still below the second peak. Finally, the model with strong magnetic
field (35OC-Rs) produces the heavy r-process elements following a
Honda-like pattern, rather than a robust, Sneden-like pattern. In
summary, MR-SNe with variations in the rotation and magnetic field can
explain the broad variability in abundance patterns found in UMP
stars. Still further investigations and models are necessary to
understand whether some of these supernovae can also produce a robust
r-process.

\section{Conclusions}
\label{sec:con}

We have presented the first nucleosynthesis yields from MR-SNe that are based on 2D simulations including detailed neutrino transport. This is critical to consistently account for the neutrino-driven ejecta and to be able to compare to the magneto-rotational ejecta. Our study is based on four models from \citet{ObergaulingerAloy2017,ObergaulingerAloy2020} with different rotation and magnetic fields. The nucleosynthesis obtained from these models can be classified into six groups depending on the abundance pattern. The model with strong magnetic field has the richest nucleosynthesis from shocked heated to r-process including also proton-rich outflows.

In all models, there is matter ejected after crossing the shock, without being accreted into the PNS, and the composition is dominated by alpha particles and iron-group elements. This matter corresponds to our $\alpha$ and $\alpha$-Fe groups. We also find a transition group with some tracers accreted down to the PNS before being ejected. These tracers correspond to the Fe-group and change from very neutron-rich when they start at the PNS surface to slightly neutron-rich or even proton-rich due to neutrinos.  Similar but a bit more neutron-rich is the Fe-weak-r-process group, where lighter heavy elements are synthesized. Finally, we find two r-process groups: the weak-r-process group that is present only in two models and the r-process group characteristic of the magneto-rotational ejecta of the model with strong magnetic fields. While the nucleosynthesis of the first groups, $\alpha$ to Fe and even Fe-weak-r-process, is typical from neutrino-driven ejecta from standard supernovae, the r-process groups are characteristics of MR-SNe.

In the model with weak magnetic field and the original progenitor rotation, there
is a late ejection of neutron-rich material at around $t\approx
1.2$~s. This enables the weak r-process to produce elements up to the second
peak. The late ejection is due to angular momentum redistribution by
the magnetic field that leads to a sudden deformation of the
neutron star. The change in the neutron star allows some neutron-rich
material to enter outflow regions. We conclude that long-time simulations of MR-SNe are critical to account for the total nucleosynthesis \citep[see also][]{Aloy_Obergaulinger_PaperII_2020arXiv200803779A}.

The model with strong magnetic field, in agreement with previous
studies \citep[e.g.,][]{Nishimura2006, Winteler2012, Saruwatari2013,
  Nishimura2015, Nishimura2017,Halevi2018, Moesta2018}, ejects
promptly neutron-rich matter. We again stress that, even more than the strength of the (poloidal) magnetic field, its large-scale, dipolar morphology is the key to produce MR-SNe \citep{Bugli2020,Aloy_Obergaulinger_PaperII_2020arXiv200803779A}. This very early ejection of matter
prevents that neutrinos change neutrons into protons and thus the
r-process successfully produces heavy elements up to the third
peak. This model (35OC-Rs) develops into a jet-like explosion with
proton-rich jets surrounded by the early-ejected, neutron-rich
material. The r-process pattern from this model does not agree with
the solar r-process. This is partially due to the uncertainties in the
nuclear physics input (see \citealt{Cowan2019,Horowitz2019} for  recent reviews) and also may indicate that the r-process in our Sun does not come only from MR-SNe but also from neutron star mergers \citep{Cote2019}. Moreover, MR-SNe were probably more frequent in the early Galaxy because low metalicity stars have lower mass-loss rates and can become fast rotators \citep{Brott2011}. 
MR-SNe can also explain the missing contribution to the europium production in the early Galaxy when one assumes only mergers as r-process site \citep{Cote2019}. We have compared the r-process pattern to observed elemental abundances
in old stars and found that our results are within the observed
abundances of what is called Honda-like stars, i.e., stars with low
enrichment of heavy r-process beyond the second peak.

We find a correlation between the explosion energy and the $^{56}$Ni
production for the three models without strong magnetic field. The
latter has higher explosion energy than the other models and relative
low $^{56}$Ni, far from what is needed to explain hypernovae. However,
our yields are lower limits as we do not consider the matter that
becomes unbound at late times or from the progenitor. Even if these
two contributions were added, we do not expect to reach the $^{56}$Ni
that is needed to explain hypernovae from the explosion. Yet, matter
ejected from the disc, which may form at late times, could provide the
missing $^{56}$Ni. In general, the amount of $^{56}$Ni and $^{44}$Ti
produced by our models is close to observed values in supernova
remnants \citep[e.g.][]{Seitenzahl2014,WangLi2016}. Moreover, we have discussed the possibility of
observing the gamma- or X-rays from the radioactive decay of heavy
elements produced in MR-SN. We have found that  $^{137}$Cs may
be observed for a MR-SN within 3~kpc.

MR-SNe show a huge
nucleosynthesis richness and may be critical to explain the early
r-process in our Galaxy. Our study demonstrates that only with MHD
simulations including detailed neutrino transport, one can accurately
calculate the complete nucleosynthesis. However, further simulations
are needed to investigate the impact of 3D and different
configurations of the magnetic field and improve our understanding of
the role of MR-SNe in the origin of heavy elements in the universe. 

\section*{Acknowledgements}
We thank Julia Bliss, Takami Kuroda, Dirk Martin,  and Friedel Thielemann for useful discussions. This research was supported by the ERC Starting Grant EUROPIUM-677912, Deutsche Forschungsgemeinschaft through SFB 1245, Helmholtz Forschungsakademie Hessen für FAIR, and by the Helmholtz-University Young Investigator grant No. VH-NG-825. MO acknowledges support from the Ministerio de Ciencia e Innovación via the Ram{\'o}n y Cajal program (RYC2018-024938-I). MAA and MO acknowledge the support by the Ministerio de Ciencia, Innovación y Universidades (PGC2018-095984-B-I00) and the Generalitat Valenciana (PROMETEU/2019/071).  This work has benefited from the COST Actions “ChETEC” (CA16117) and "PHAROS” (CA16214), supported by COST (European Cooperation in Science and Technology). Nucleosynthesis computations were performed on the Lichtenberg High Performance Computer (TU Darmstadt). MAA and MO thankfully acknowledge the computer resources and the
technical support provided by the grants AECT-2017-2-0006,
AECT-2017-3-0007, AECT-2018-1-0010, AECT-2018-2-0003,
AECT-2018-3-0010, and AECT-2019-1-0009 of the Spanish Supercomputing
Network on clusters \textit{Pirineus} of the Consorci de Serveis
Universitaris de Catalunya (CSUC), \textit{Picasso} of the Universidad
de M{\'a}laga, and \textit{MareNostrum} of the Barcelona
Supercomputing Centre - Centro Nacional de Supercomputaci\'on,
respectively, and on the clusters \textit{Tirant} and
\textit{Lluisvives} of the Servei d'Inform\`atica of the University of
Valencia (financed by the FEDER funds for Scientific Infrastructures;
IDIFEDER-2018-063).

\section*{Data Availability}
Ten representative tracer particles of every nucleosynthesis group of Model 35OC-Rs can be found at \url{https://theorie.ikp.physik.tu-darmstadt.de/astro/resources.php} and \url{https://github.com/nuc-astro/RepresentativeTrajectories_MagnetoRotationalSupernova}. The integrated nucleosynthetic yields of the innermost part of the ejecta are given in the appendix.

\bibliographystyle{mnras}
\bibliography{paper} 



\appendix
\section{Data tables}
\label{app:datatables}
The  mass fractions after 1 Gyr can be found in Tables~\ref{tab:data_ro}, \ref{tab:data_rw}, \ref{tab:data_rrw}, and \ref{tab:data_rs}. We want to stress that, for the light elements, we provide only the contribution from the innermost part of the ejecta. The progenitor composition  and late nucleosynthesis should be considered if supernova yields are needed for the lighter elements. All tables are also available online.
\begin{table*}
\caption{Mass fractions of individual isotopes for model 35OC-RO after 1 Gyr. The yields represent only the innermost part of the ejecta.}
\label{tab:data_ro}
 \centering
\begin{tabular}{cr|cr|cr|cr|cr}
\hline
       Isotope & $X_i$& Isotope & $X_i$&Isotope & $X_i$&Isotope & $X_i$&Isotope & $X_i$ \\
\hline
       $^{1}$H &  $2.74\cdot 10^{-02}$ &  $^{28}$Si &  $3.88\cdot 10^{-02}$ &  $^{48}$Ti &  $1.29\cdot 10^{-03}$  & $^{68}$Zn &  $2.48\cdot 10^{-03}$  & $^{85}$Rb &  $3.34\cdot 10^{-05}$     \\
    $^{2}$H &  $6.95\cdot 10^{-09}$ &  $^{29}$Si &  $5.63\cdot 10^{-05}$ &  $^{49}$Ti &  $1.37\cdot 10^{-04}$  & $^{70}$Zn &  $5.44\cdot 10^{-05}$  & $^{87}$Rb &  $4.47\cdot 10^{-05}$     \\
   $^{3}$He &  $3.97\cdot 10^{-08}$ &  $^{30}$Si &  $3.21\cdot 10^{-04}$ & $^{50}$Ti &  $3.01\cdot 10^{-04}$  &  $^{69}$Ga &  $7.98\cdot 10^{-05}$  & $^{84}$Sr &  $2.05\cdot 10^{-06}$     \\
   $^{4}$He &  $1.46\cdot 10^{-01}$ &   $^{31}$P &  $1.07\cdot 10^{-04}$ &  $^{51}$V &  $1.17\cdot 10^{-04}$  &   $^{71}$Ga &  $1.77\cdot 10^{-05}$  &$^{86}$Sr &  $5.96\cdot 10^{-06}$      \\
   $^{7}$Li &  $2.60\cdot 10^{-09}$ &   $^{32}$S &  $1.93\cdot 10^{-02}$ & $^{50}$Cr &  $1.06\cdot 10^{-04}$  &   $^{70}$Ge &  $1.01\cdot 10^{-03}$  & $^{87}$Sr &  $2.17\cdot 10^{-06}$     \\
   $^{7}$Be &  $4.97\cdot 10^{-08}$ &   $^{33}$S &  $7.41\cdot 10^{-05}$ & $^{52}$Cr &  $9.11\cdot 10^{-04}$  &  $^{72}$Ge &  $2.26\cdot 10^{-04}$   & $^{88}$Sr &  $6.74\cdot 10^{-04}$     \\
   $^{11}$B &  $1.51\cdot 10^{-09}$ &   $^{34}$S &  $3.90\cdot 10^{-03}$ & $^{53}$Cr &  $1.58\cdot 10^{-04}$  &  $^{73}$Ge &  $1.04\cdot 10^{-05}$   &  $^{89}$Y &  $6.93\cdot 10^{-05}$     \\
   $^{12}$C &  $9.27\cdot 10^{-03}$ &   $^{36}$S &  $3.31\cdot 10^{-08}$ & $^{54}$Cr &  $7.43\cdot 10^{-04}$  &  $^{74}$Ge &  $1.01\cdot 10^{-04}$   &  $^{90}$Zr &  $6.22\cdot 10^{-04}$    \\
   $^{13}$C &  $4.64\cdot 10^{-08}$ &  $^{35}$Cl &  $1.12\cdot 10^{-04}$ & $^{55}$Mn &  $4.76\cdot 10^{-04}$  &  $^{76}$Ge &  $4.32\cdot 10^{-05}$   &  $^{91}$Zr &  $4.77\cdot 10^{-06}$    \\
   $^{14}$N &  $7.43\cdot 10^{-07}$ &  $^{37}$Cl &  $6.02\cdot 10^{-05}$ & $^{54}$Fe &  $2.36\cdot 10^{-03}$  &  $^{75}$As &  $1.14\cdot 10^{-05}$   &  $^{92}$Zr &  $6.32\cdot 10^{-08}$    \\
   $^{15}$N &  $3.11\cdot 10^{-06}$ &  $^{36}$Ar &  $3.64\cdot 10^{-03}$ & $^{56}$Fe &  $1.47\cdot 10^{-01}$  &  $^{74}$Se &  $7.32\cdot 10^{-05}$   &  $^{93}$Nb &  $1.09\cdot 10^{-07}$    \\
   $^{16}$O &  $3.40\cdot 10^{-01}$ &  $^{38}$Ar &  $2.40\cdot 10^{-03}$ & $^{57}$Fe &  $4.18\cdot 10^{-03}$  &  $^{76}$Se &  $1.09\cdot 10^{-04}$   &  $^{92}$Mo &  $7.16\cdot 10^{-06}$    \\
   $^{17}$O &  $1.34\cdot 10^{-06}$ &  $^{40}$Ar &  $5.72\cdot 10^{-09}$ & $^{58}$Fe &  $2.78\cdot 10^{-03}$  &  $^{77}$Se &  $1.19\cdot 10^{-05}$   &  $^{94}$Mo &  $6.01\cdot 10^{-08}$    \\
   $^{18}$O &  $3.14\cdot 10^{-08}$ &   $^{39}$K &  $1.83\cdot 10^{-04}$ & $^{59}$Co &  $1.97\cdot 10^{-03}$  &  $^{78}$Se &  $1.23\cdot 10^{-04}$   &  $^{95}$Mo &  $1.55\cdot 10^{-09}$    \\
   $^{19}$F &  $2.62\cdot 10^{-09}$ &   $^{40}$K &  $1.49\cdot 10^{-08}$ & $^{58}$Ni &  $3.70\cdot 10^{-02}$  &  $^{80}$Se &  $2.33\cdot 10^{-04}$   &  $^{97}$Mo &  $1.26\cdot 10^{-09}$    \\
  $^{20}$Ne &  $6.80\cdot 10^{-02}$ &   $^{41}$K &  $2.14\cdot 10^{-05}$ & $^{60}$Ni &  $4.90\cdot 10^{-02}$  &  $^{82}$Se &  $1.04\cdot 10^{-04}$   &  $^{96}$Ru &  $9.43\cdot 10^{-10}$    \\
  $^{21}$Ne &  $1.17\cdot 10^{-07}$ &  $^{40}$Ca &  $3.37\cdot 10^{-03}$ & $^{61}$Ni &  $1.18\cdot 10^{-03}$  &  $^{79}$Br &  $1.11\cdot 10^{-05}$   &  $^{98}$Ru &  $3.26\cdot 10^{-09}$    \\
  $^{22}$Ne &  $2.14\cdot 10^{-06}$ &  $^{42}$Ca &  $2.53\cdot 10^{-04}$ & $^{62}$Ni &  $2.96\cdot 10^{-02}$  &  $^{81}$Br &  $2.90\cdot 10^{-05}$   &  $^{99}$Ru &  $4.84\cdot 10^{-10}$    \\
  $^{23}$Na &  $8.86\cdot 10^{-06}$ &  $^{43}$Ca &  $3.86\cdot 10^{-05}$ & $^{64}$Ni &  $9.14\cdot 10^{-03}$  &  $^{78}$Kr &  $7.84\cdot 10^{-06}$   & $^{100}$Ru &  $3.94\cdot 10^{-10}$\\
  $^{24}$Mg &  $1.22\cdot 10^{-02}$ &  $^{44}$Ca &  $2.07\cdot 10^{-04}$ & $^{63}$Cu &  $1.13\cdot 10^{-03}$  &  $^{80}$Kr &  $2.57\cdot 10^{-05}$   & $^{101}$Ru &  $3.89\cdot 10^{-10}$\\
  $^{25}$Mg &  $1.19\cdot 10^{-06}$ &  $^{46}$Ca &  $1.48\cdot 10^{-09}$ & $^{65}$Cu &  $3.83\cdot 10^{-04}$  &  $^{82}$Kr &  $1.61\cdot 10^{-05}$   & $^{103}$Rh &  $1.92\cdot 10^{-10}$\\
  $^{26}$Mg &  $4.50\cdot 10^{-08}$ &  $^{45}$Sc &  $4.70\cdot 10^{-05}$ & $^{64}$Zn &  $1.46\cdot 10^{-02}$  &  $^{83}$Kr &  $5.04\cdot 10^{-05}$   & $^{102}$Pd &  $5.34\cdot 10^{-10}$\\
  $^{26}$Al &  $7.07\cdot 10^{-07}$ &  $^{46}$Ti &  $4.38\cdot 10^{-05}$ & $^{66}$Zn &  $1.32\cdot 10^{-02}$  &  $^{84}$Kr &  $2.91\cdot 10^{-04}$   & \\
  $^{27}$Al &  $1.56\cdot 10^{-05}$ &  $^{47}$Ti &  $6.67\cdot 10^{-05}$ & $^{67}$Zn &  $1.45\cdot 10^{-04}$  &  $^{86}$Kr &  $3.11\cdot 10^{-04}$   & \\
  \hline
    \multicolumn{10}{l}{The table is also published in machine-readable format.}\\
\end{tabular}

\end{table*}

\begin{table*}
\caption{Mass fractions of individual isotopes for model 35OC-Rw after 1 Gyr. The yields represent only the innermost part of the ejecta. }
\label{tab:data_rw}
 \centering
\begin{tabular}{cr|cr|cr|cr|cr}
\hline
       Isotope & $X_i$& Isotope & $X_i$&Isotope & $X_i$&Isotope & $X_i$&Isotope & $X_i$ \\
\hline
    $^{1}$H &  $1.19\cdot 10^{-02}$ &  $^{37}$Cl &  $5.63\cdot 10^{-05}$ &  $^{64}$Ni &  $1.96\cdot 10^{-03}$ &  $^{86}$Sr &  $3.55\cdot 10^{-06}$ &   $^{112}$Cd &  $1.11\cdot 10^{-05}$\\
    $^{2}$H &  $2.93\cdot 10^{-07}$ &  $^{36}$Ar &  $4.00\cdot 10^{-03}$ &  $^{63}$Cu &  $4.84\cdot 10^{-04}$ &  $^{87}$Sr &  $2.14\cdot 10^{-06}$ &   $^{113}$Cd &  $2.11\cdot 10^{-05}$\\
   $^{3}$He &  $6.29\cdot 10^{-08}$ &  $^{38}$Ar &  $2.49\cdot 10^{-03}$ &  $^{65}$Cu &  $1.50\cdot 10^{-04}$ &  $^{88}$Sr &  $6.48\cdot 10^{-04}$ &   $^{114}$Cd &  $8.16\cdot 10^{-06}$\\
   $^{4}$He &  $1.24\cdot 10^{-01}$ &  $^{40}$Ar &  $2.38\cdot 10^{-06}$ &  $^{64}$Zn &  $4.68\cdot 10^{-03}$ &   $^{89}$Y &  $1.04\cdot 10^{-04}$ &   $^{116}$Cd &  $1.57\cdot 10^{-05}$\\
   $^{7}$Li &  $4.21\cdot 10^{-09}$ &   $^{39}$K &  $1.52\cdot 10^{-04}$ &  $^{66}$Zn &  $4.41\cdot 10^{-03}$ &  $^{90}$Zr &  $6.43\cdot 10^{-04}$ &   $^{115}$In &  $1.28\cdot 10^{-06}$\\
   $^{7}$Be &  $2.34\cdot 10^{-08}$ &   $^{40}$K &  $1.61\cdot 10^{-05}$ &  $^{67}$Zn &  $9.22\cdot 10^{-05}$ &  $^{91}$Zr &  $5.59\cdot 10^{-05}$ &   $^{114}$Sn &  $9.85\cdot 10^{-10}$\\
   $^{11}$B &  $3.02\cdot 10^{-06}$ &   $^{41}$K &  $2.80\cdot 10^{-05}$ &  $^{68}$Zn &  $1.42\cdot 10^{-03}$ &  $^{92}$Zr &  $8.49\cdot 10^{-06}$ &   $^{116}$Sn &  $1.44\cdot 10^{-09}$\\
   $^{12}$C &  $6.73\cdot 10^{-03}$ &  $^{40}$Ca &  $4.08\cdot 10^{-03}$ &  $^{70}$Zn &  $2.30\cdot 10^{-04}$ &  $^{94}$Zr &  $2.00\cdot 10^{-05}$ &   $^{117}$Sn &  $2.48\cdot 10^{-06}$\\
   $^{13}$C &  $4.92\cdot 10^{-07}$ &  $^{42}$Ca &  $2.52\cdot 10^{-04}$ &  $^{69}$Ga &  $1.33\cdot 10^{-04}$ &  $^{93}$Nb &  $6.78\cdot 10^{-06}$ &   $^{118}$Sn &  $4.11\cdot 10^{-06}$\\
   $^{14}$N &  $4.04\cdot 10^{-06}$ &  $^{43}$Ca &  $2.72\cdot 10^{-05}$ &  $^{71}$Ga &  $1.03\cdot 10^{-04}$ &  $^{92}$Mo &  $3.95\cdot 10^{-05}$ &   $^{119}$Sn &  $1.19\cdot 10^{-06}$\\
   $^{15}$N &  $2.55\cdot 10^{-05}$ &  $^{44}$Ca &  $2.61\cdot 10^{-04}$ &  $^{70}$Ge &  $5.36\cdot 10^{-04}$ &  $^{94}$Mo &  $1.76\cdot 10^{-07}$ &   $^{120}$Sn &  $2.06\cdot 10^{-06}$\\
   $^{16}$O &  $2.57\cdot 10^{-01}$ &  $^{46}$Ca &  $3.85\cdot 10^{-07}$ &  $^{72}$Ge &  $2.26\cdot 10^{-04}$ &  $^{95}$Mo &  $8.74\cdot 10^{-06}$ &   $^{122}$Sn &  $1.97\cdot 10^{-06}$\\
   $^{17}$O &  $1.35\cdot 10^{-05}$ &  $^{45}$Sc &  $2.33\cdot 10^{-05}$ &  $^{73}$Ge &  $1.49\cdot 10^{-04}$ &  $^{96}$Mo &  $9.07\cdot 10^{-06}$ &   $^{124}$Sn &  $1.70\cdot 10^{-06}$\\
   $^{18}$O &  $2.39\cdot 10^{-08}$ &  $^{46}$Ti &  $2.63\cdot 10^{-05}$ &  $^{74}$Ge &  $2.85\cdot 10^{-04}$ &  $^{97}$Mo &  $1.25\cdot 10^{-06}$ &   $^{121}$Sb &  $7.15\cdot 10^{-07}$\\
   $^{19}$F &  $1.13\cdot 10^{-05}$ &  $^{47}$Ti &  $4.72\cdot 10^{-05}$ &  $^{76}$Ge &  $5.96\cdot 10^{-04}$ &  $^{98}$Mo &  $5.54\cdot 10^{-06}$ &   $^{123}$Sb &  $1.38\cdot 10^{-06}$\\
  $^{20}$Ne &  $4.90\cdot 10^{-02}$ &  $^{48}$Ti &  $4.10\cdot 10^{-03}$ &  $^{75}$As &  $9.92\cdot 10^{-05}$ & $^{100}$Mo &  $1.44\cdot 10^{-05}$ &   $^{122}$Te &  $1.48\cdot 10^{-09}$\\
  $^{21}$Ne &  $2.80\cdot 10^{-05}$ &  $^{49}$Ti &  $1.39\cdot 10^{-04}$ &  $^{74}$Se &  $4.87\cdot 10^{-05}$ &  $^{96}$Ru &  $2.93\cdot 10^{-09}$ &   $^{124}$Te &  $1.86\cdot 10^{-09}$\\
  $^{22}$Ne &  $1.42\cdot 10^{-06}$ &  $^{50}$Ti &  $1.24\cdot 10^{-03}$ &  $^{76}$Se &  $5.98\cdot 10^{-05}$ &  $^{98}$Ru &  $1.45\cdot 10^{-09}$ &   $^{125}$Te &  $3.48\cdot 10^{-06}$\\
  $^{23}$Na &  $2.44\cdot 10^{-05}$ &   $^{51}$V &  $4.46\cdot 10^{-04}$ &  $^{77}$Se &  $1.13\cdot 10^{-04}$ &  $^{99}$Ru &  $1.20\cdot 10^{-06}$ &   $^{126}$Te &  $1.43\cdot 10^{-06}$\\
  $^{24}$Mg &  $9.91\cdot 10^{-03}$ &  $^{50}$Cr &  $5.65\cdot 10^{-05}$ &  $^{78}$Se &  $3.04\cdot 10^{-04}$ & $^{100}$Ru &  $8.66\cdot 10^{-09}$ &   $^{128}$Te &  $4.06\cdot 10^{-06}$\\
  $^{25}$Mg &  $4.93\cdot 10^{-05}$ &  $^{52}$Cr &  $1.71\cdot 10^{-03}$ &  $^{80}$Se &  $1.81\cdot 10^{-03}$ & $^{101}$Ru &  $1.76\cdot 10^{-06}$ &   $^{130}$Te &  $2.02\cdot 10^{-05}$\\
  $^{26}$Mg &  $3.07\cdot 10^{-06}$ &  $^{53}$Cr &  $8.76\cdot 10^{-04}$ &  $^{82}$Se &  $4.67\cdot 10^{-03}$ & $^{102}$Ru &  $7.38\cdot 10^{-06}$ &    $^{127}$I &  $2.35\cdot 10^{-06}$\\
  $^{26}$Al &  $5.16\cdot 10^{-06}$ &  $^{54}$Cr &  $1.34\cdot 10^{-03}$ &  $^{79}$Br &  $7.09\cdot 10^{-04}$ & $^{104}$Ru &  $1.85\cdot 10^{-05}$ &   $^{129}$Xe &  $4.66\cdot 10^{-06}$\\
  $^{27}$Al &  $2.07\cdot 10^{-04}$ &  $^{55}$Mn &  $1.59\cdot 10^{-03}$ &  $^{81}$Br &  $7.95\cdot 10^{-04}$ & $^{103}$Rh &  $8.95\cdot 10^{-06}$ &   $^{130}$Xe &  $2.39\cdot 10^{-10}$\\
  $^{28}$Si &  $3.87\cdot 10^{-02}$ &  $^{54}$Fe &  $2.94\cdot 10^{-03}$ &  $^{78}$Kr &  $5.92\cdot 10^{-06}$ & $^{104}$Pd &  $3.53\cdot 10^{-10}$ &   $^{131}$Xe &  $4.96\cdot 10^{-06}$\\
  $^{29}$Si &  $1.30\cdot 10^{-04}$ &  $^{56}$Fe &  $3.22\cdot 10^{-01}$ &  $^{80}$Kr &  $1.37\cdot 10^{-05}$ & $^{105}$Pd &  $9.01\cdot 10^{-06}$ &   $^{132}$Xe &  $5.13\cdot 10^{-04}$\\
  $^{30}$Si &  $3.36\cdot 10^{-04}$ &  $^{57}$Fe &  $1.53\cdot 10^{-02}$ &  $^{82}$Kr &  $1.08\cdot 10^{-05}$ & $^{106}$Pd &  $2.80\cdot 10^{-05}$ &   $^{134}$Xe &  $4.20\cdot 10^{-07}$\\
   $^{31}$P &  $1.97\cdot 10^{-04}$ &  $^{58}$Fe &  $1.29\cdot 10^{-03}$ &  $^{83}$Kr &  $4.90\cdot 10^{-03}$ & $^{108}$Pd &  $1.14\cdot 10^{-06}$ &   $^{133}$Cs &  $3.37\cdot 10^{-05}$\\
   $^{32}$S &  $2.03\cdot 10^{-02}$ &  $^{59}$Co &  $1.80\cdot 10^{-03}$ &  $^{84}$Kr &  $2.01\cdot 10^{-03}$ & $^{110}$Pd &  $4.43\cdot 10^{-06}$ &   $^{135}$Ba &  $7.26\cdot 10^{-10}$\\
   $^{33}$S &  $1.51\cdot 10^{-04}$ &  $^{58}$Ni &  $4.14\cdot 10^{-02}$ &  $^{86}$Kr &  $1.90\cdot 10^{-04}$ & $^{107}$Ag &  $6.46\cdot 10^{-06}$ & \\
   $^{34}$S &  $4.10\cdot 10^{-03}$ &  $^{60}$Ni &  $2.39\cdot 10^{-02}$ &  $^{85}$Rb &  $1.99\cdot 10^{-04}$ & $^{109}$Ag &  $7.94\cdot 10^{-07}$ & \\
   $^{36}$S &  $2.49\cdot 10^{-06}$ &  $^{61}$Ni &  $9.04\cdot 10^{-04}$ &  $^{87}$Rb &  $6.57\cdot 10^{-05}$ & $^{110}$Cd &  $2.44\cdot 10^{-09}$ & \\
  $^{35}$Cl &  $1.48\cdot 10^{-04}$ &  $^{62}$Ni &  $1.12\cdot 10^{-02}$ &  $^{84}$Sr &  $1.44\cdot 10^{-06}$ & $^{111}$Cd &  $1.78\cdot 10^{-06}$ & \\
  \hline
    \multicolumn{10}{l}{The table is also published in machine-readable format.}
\end{tabular}
\end{table*}

\begin{table*}
\caption{Mass fractions of individual isotopes for model 35OC-RRw after 1 Gyr. The yields represent only the innermost part of the ejecta.}
\label{tab:data_rrw}
 \centering
\begin{tabular}{cr|cr|cr|cr|cr}
\hline
       Isotope & $X_i$& Isotope & $X_i$&Isotope & $X_i$&Isotope & $X_i$&Isotope & $X_i$ \\
\hline
   $^{1}$H &  $3.40\cdot 10^{-02}$ & $^{27}$Al &  $1.41\cdot 10^{-03}$ &  $^{46}$Ti &  $1.24\cdot 10^{-04}$ &  $^{64}$Zn &  $6.56\cdot 10^{-03}$ & $^{82}$Kr &  $5.56\cdot 10^{-06}$ \\
   $^{2}$H &  $7.76\cdot 10^{-07}$ & $^{28}$Si &  $1.51\cdot 10^{-01}$ &  $^{47}$Ti &  $1.17\cdot 10^{-04}$ &  $^{66}$Zn &  $1.68\cdot 10^{-04}$ & $^{83}$Kr &  $4.08\cdot 10^{-06}$ \\
  $^{3}$He &  $4.95\cdot 10^{-07}$ & $^{29}$Si &  $8.14\cdot 10^{-04}$ &  $^{48}$Ti &  $3.28\cdot 10^{-04}$ &  $^{67}$Zn &  $2.13\cdot 10^{-04}$ & $^{84}$Kr &  $1.72\cdot 10^{-07}$ \\
  $^{4}$He &  $2.11\cdot 10^{-01}$ & $^{30}$Si &  $2.57\cdot 10^{-03}$ &  $^{49}$Ti &  $1.81\cdot 10^{-04}$ &  $^{68}$Zn &  $1.21\cdot 10^{-03}$ & $^{86}$Kr &  $1.23\cdot 10^{-08}$ \\
  $^{7}$Li &  $3.52\cdot 10^{-08}$ &  $^{31}$P &  $1.78\cdot 10^{-03}$ &  $^{50}$Ti &  $6.16\cdot 10^{-06}$ &  $^{70}$Zn &  $6.44\cdot 10^{-07}$ & $^{85}$Rb &  $2.58\cdot 10^{-06}$ \\
  $^{7}$Be &  $1.24\cdot 10^{-07}$ &  $^{32}$S &  $5.96\cdot 10^{-02}$ &   $^{51}$V &  $2.14\cdot 10^{-04}$ &  $^{69}$Ga &  $2.04\cdot 10^{-04}$ & $^{87}$Rb &  $1.61\cdot 10^{-08}$ \\
  $^{11}$B &  $7.45\cdot 10^{-07}$ &  $^{33}$S &  $1.01\cdot 10^{-03}$ &  $^{50}$Cr &  $1.94\cdot 10^{-04}$ &  $^{71}$Ga &  $4.62\cdot 10^{-05}$ & $^{84}$Sr &  $2.62\cdot 10^{-06}$ \\
  $^{12}$C &  $2.87\cdot 10^{-04}$ &  $^{34}$S &  $3.79\cdot 10^{-02}$ &  $^{52}$Cr &  $1.61\cdot 10^{-03}$ &  $^{70}$Ge &  $5.51\cdot 10^{-05}$ & $^{86}$Sr &  $1.10\cdot 10^{-06}$ \\
  $^{13}$C &  $4.97\cdot 10^{-06}$ &  $^{36}$S &  $5.95\cdot 10^{-05}$ &  $^{53}$Cr &  $3.91\cdot 10^{-04}$ &  $^{72}$Ge &  $2.14\cdot 10^{-04}$ & $^{87}$Sr &  $6.48\cdot 10^{-07}$ \\
  $^{14}$N &  $4.62\cdot 10^{-06}$ & $^{35}$Cl &  $2.53\cdot 10^{-03}$ &  $^{54}$Cr &  $2.86\cdot 10^{-04}$ &  $^{73}$Ge &  $5.72\cdot 10^{-05}$ & $^{88}$Sr &  $4.33\cdot 10^{-07}$ \\
  $^{15}$N &  $6.38\cdot 10^{-05}$ & $^{37}$Cl &  $2.84\cdot 10^{-04}$ &  $^{55}$Mn &  $1.99\cdot 10^{-03}$ &  $^{74}$Ge &  $3.16\cdot 10^{-06}$ &  $^{89}$Y &  $2.44\cdot 10^{-07}$ \\
  $^{16}$O &  $1.62\cdot 10^{-01}$ & $^{36}$Ar &  $1.14\cdot 10^{-02}$ &  $^{54}$Fe &  $3.04\cdot 10^{-03}$ &  $^{76}$Ge &  $2.26\cdot 10^{-07}$ & $^{90}$Zr &  $3.99\cdot 10^{-06}$ \\
  $^{17}$O &  $1.80\cdot 10^{-05}$ & $^{38}$Ar &  $2.15\cdot 10^{-02}$ &  $^{56}$Fe &  $2.13\cdot 10^{-01}$ &  $^{75}$As &  $1.60\cdot 10^{-05}$ & $^{91}$Zr &  $2.12\cdot 10^{-06}$ \\
  $^{18}$O &  $4.83\cdot 10^{-08}$ & $^{40}$Ar &  $5.76\cdot 10^{-05}$ &  $^{57}$Fe &  $6.21\cdot 10^{-03}$ &  $^{74}$Se &  $2.27\cdot 10^{-05}$ & $^{92}$Zr &  $6.27\cdot 10^{-08}$ \\
  $^{19}$F &  $8.74\cdot 10^{-06}$ &  $^{39}$K &  $1.72\cdot 10^{-03}$ &  $^{58}$Fe &  $2.22\cdot 10^{-04}$ &  $^{76}$Se &  $4.30\cdot 10^{-05}$ & $^{93}$Nb &  $2.85\cdot 10^{-06}$ \\
 $^{20}$Ne &  $2.49\cdot 10^{-03}$ &  $^{40}$K &  $2.81\cdot 10^{-04}$ &  $^{59}$Co &  $1.71\cdot 10^{-03}$ &  $^{77}$Se &  $2.33\cdot 10^{-05}$ & $^{92}$Mo &  $1.21\cdot 10^{-04}$ \\
 $^{21}$Ne &  $4.28\cdot 10^{-06}$ &  $^{41}$K &  $1.26\cdot 10^{-04}$ &  $^{58}$Ni &  $4.29\cdot 10^{-03}$ &  $^{78}$Se &  $6.72\cdot 10^{-06}$ & $^{94}$Mo &  $4.93\cdot 10^{-07}$ \\
 $^{22}$Ne &  $2.23\cdot 10^{-06}$ & $^{40}$Ca &  $7.47\cdot 10^{-03}$ &  $^{60}$Ni &  $2.83\cdot 10^{-02}$ &  $^{80}$Se &  $2.63\cdot 10^{-07}$ & $^{95}$Mo &  $7.30\cdot 10^{-08}$ \\
 $^{23}$Na &  $4.05\cdot 10^{-05}$ & $^{42}$Ca &  $3.49\cdot 10^{-03}$ &  $^{61}$Ni &  $6.63\cdot 10^{-04}$ &  $^{82}$Se &  $1.39\cdot 10^{-08}$ & $^{96}$Mo &  $1.07\cdot 10^{-09}$ \\
 $^{24}$Mg &  $1.01\cdot 10^{-02}$ & $^{43}$Ca &  $1.04\cdot 10^{-04}$ &  $^{62}$Ni &  $8.90\cdot 10^{-04}$ &  $^{79}$Br &  $7.58\cdot 10^{-06}$ & $^{97}$Mo &  $1.62\cdot 10^{-09}$ \\
 $^{25}$Mg &  $1.45\cdot 10^{-04}$ & $^{44}$Ca &  $3.59\cdot 10^{-04}$ &  $^{64}$Ni &  $3.25\cdot 10^{-05}$ &  $^{81}$Br &  $9.38\cdot 10^{-06}$ & $^{96}$Ru &  $3.97\cdot 10^{-08}$ \\
 $^{26}$Mg &  $1.05\cdot 10^{-05}$ & $^{46}$Ca &  $9.06\cdot 10^{-08}$ &  $^{63}$Cu &  $1.02\cdot 10^{-03}$ &  $^{78}$Kr &  $9.56\cdot 10^{-06}$ & $^{98}$Ru &  $2.44\cdot 10^{-10}$ \\
 $^{26}$Al &  $1.23\cdot 10^{-05}$ & $^{45}$Sc &  $9.51\cdot 10^{-05}$ &  $^{65}$Cu &  $4.26\cdot 10^{-04}$ &  $^{80}$Kr &  $9.27\cdot 10^{-06}$ & \\
 \hline
   \multicolumn{10}{l}{The table is also published in machine-readable format.}                     
\end{tabular}

\end{table*}

\begin{table*}
\caption{Mass fractions of individual isotopes for model 35OC-Rs after 1 Gyr. The yields represent only the innermost part of the ejecta.}
\label{tab:data_rs}
 \centering
\begin{tabular}{cr|cr|cr|cr|cr}
\hline
       Isotope & $X_i$& Isotope & $X_i$&Isotope & $X_i$&Isotope & $X_i$&Isotope & $X_i$ \\
\hline
    $^{1}$H &  $9.57\cdot 10^{-04}$ &  $^{50}$Ti &  $3.06\cdot 10^{-02}$ &    $^{85}$Rb &  $1.15\cdot 10^{-03}$ &   $^{126}$Te &  $3.39\cdot 10^{-04}$ & $^{172}$Yb &  $1.75\cdot 10^{-05}$  \\
   $^{4}$He &  $2.64\cdot 10^{-02}$ &   $^{51}$V &  $4.17\cdot 10^{-03}$ &    $^{87}$Rb &  $4.82\cdot 10^{-04}$ &   $^{128}$Te &  $8.52\cdot 10^{-04}$ & $^{173}$Yb &  $2.43\cdot 10^{-05}$  \\
   $^{7}$Be &  $1.36\cdot 10^{-10}$ &  $^{50}$Cr &  $5.60\cdot 10^{-05}$ &    $^{84}$Sr &  $4.24\cdot 10^{-07}$ &   $^{130}$Te &  $3.49\cdot 10^{-03}$ & $^{174}$Yb &  $2.88\cdot 10^{-05}$  \\
   $^{12}$C &  $4.79\cdot 10^{-03}$ &  $^{52}$Cr &  $5.13\cdot 10^{-03}$ &    $^{86}$Sr &  $2.42\cdot 10^{-06}$ &    $^{127}$I &  $3.17\cdot 10^{-04}$ & $^{176}$Yb &  $1.04\cdot 10^{-05}$  \\
   $^{13}$C &  $5.94\cdot 10^{-10}$ &  $^{53}$Cr &  $1.65\cdot 10^{-02}$ &    $^{87}$Sr &  $8.25\cdot 10^{-06}$ &   $^{129}$Xe &  $1.86\cdot 10^{-03}$ & $^{175}$Lu &  $9.23\cdot 10^{-06}$  \\
   $^{14}$N &  $2.03\cdot 10^{-07}$ &  $^{54}$Cr &  $9.32\cdot 10^{-03}$ &    $^{88}$Sr &  $2.65\cdot 10^{-03}$ &   $^{131}$Xe &  $3.29\cdot 10^{-03}$ & $^{177}$Hf &  $5.85\cdot 10^{-06}$  \\
   $^{15}$N &  $4.48\cdot 10^{-07}$ &  $^{55}$Mn &  $7.04\cdot 10^{-04}$ &     $^{89}$Y &  $5.74\cdot 10^{-04}$ &   $^{132}$Xe &  $5.02\cdot 10^{-03}$ & $^{178}$Hf &  $1.37\cdot 10^{-05}$  \\
   $^{16}$O &  $2.46\cdot 10^{-01}$ &  $^{54}$Fe &  $3.92\cdot 10^{-03}$ &    $^{90}$Zr &  $5.44\cdot 10^{-04}$ &   $^{134}$Xe &  $2.58\cdot 10^{-04}$ & $^{179}$Hf &  $9.21\cdot 10^{-06}$  \\
   $^{17}$O &  $4.56\cdot 10^{-08}$ &  $^{56}$Fe &  $7.01\cdot 10^{-02}$ &    $^{91}$Zr &  $1.17\cdot 10^{-04}$ &   $^{136}$Xe &  $2.32\cdot 10^{-05}$ & $^{180}$Hf &  $1.19\cdot 10^{-05}$  \\
   $^{18}$O &  $3.25\cdot 10^{-06}$ &  $^{57}$Fe &  $5.13\cdot 10^{-03}$ &    $^{92}$Zr &  $7.59\cdot 10^{-05}$ &   $^{133}$Cs &  $8.67\cdot 10^{-04}$ & $^{181}$Ta &  $5.21\cdot 10^{-06}$  \\
   $^{19}$F &  $3.07\cdot 10^{-06}$ &  $^{58}$Fe &  $1.49\cdot 10^{-02}$ &    $^{94}$Zr &  $1.49\cdot 10^{-04}$ &   $^{135}$Ba &  $1.32\cdot 10^{-05}$ &  $^{182}$W &  $8.61\cdot 10^{-06}$  \\
  $^{20}$Ne &  $3.48\cdot 10^{-02}$ &  $^{59}$Co &  $1.35\cdot 10^{-03}$ &    $^{93}$Nb &  $2.81\cdot 10^{-04}$ &   $^{137}$Ba &  $8.25\cdot 10^{-06}$ &  $^{183}$W &  $1.32\cdot 10^{-06}$  \\
  $^{21}$Ne &  $6.38\cdot 10^{-07}$ &  $^{58}$Ni &  $4.87\cdot 10^{-02}$ &    $^{92}$Mo &  $1.86\cdot 10^{-06}$ &   $^{138}$Ba &  $3.19\cdot 10^{-05}$ &  $^{184}$W &  $3.75\cdot 10^{-06}$  \\
  $^{22}$Ne &  $3.49\cdot 10^{-06}$ &  $^{60}$Ni &  $2.93\cdot 10^{-02}$ &    $^{94}$Mo &  $2.15\cdot 10^{-08}$ &   $^{139}$La &  $7.63\cdot 10^{-06}$ &  $^{186}$W &  $1.23\cdot 10^{-06}$  \\
  $^{23}$Na &  $9.55\cdot 10^{-06}$ &  $^{61}$Ni &  $1.18\cdot 10^{-03}$ &    $^{95}$Mo &  $1.44\cdot 10^{-04}$ &   $^{140}$Ce &  $2.16\cdot 10^{-06}$ & $^{185}$Re &  $8.96\cdot 10^{-07}$  \\
  $^{24}$Mg &  $1.17\cdot 10^{-02}$ &  $^{62}$Ni &  $4.59\cdot 10^{-02}$ &    $^{96}$Mo &  $5.93\cdot 10^{-05}$ &   $^{142}$Ce &  $3.09\cdot 10^{-06}$ & $^{187}$Re &  $3.90\cdot 10^{-07}$  \\
  $^{25}$Mg &  $8.32\cdot 10^{-07}$ &  $^{64}$Ni &  $3.23\cdot 10^{-02}$ &    $^{97}$Mo &  $4.43\cdot 10^{-05}$ &   $^{141}$Pr &  $2.04\cdot 10^{-06}$ & $^{187}$Os &  $6.27\cdot 10^{-09}$  \\
  $^{26}$Mg &  $6.10\cdot 10^{-07}$ &  $^{63}$Cu &  $1.38\cdot 10^{-03}$ &    $^{98}$Mo &  $4.70\cdot 10^{-05}$ &   $^{143}$Nd &  $3.89\cdot 10^{-06}$ & $^{188}$Os &  $1.26\cdot 10^{-06}$  \\
  $^{26}$Al &  $9.52\cdot 10^{-07}$ &  $^{65}$Cu &  $6.82\cdot 10^{-04}$ &   $^{100}$Mo &  $6.90\cdot 10^{-05}$ &   $^{144}$Nd &  $2.80\cdot 10^{-06}$ & $^{189}$Os &  $9.44\cdot 10^{-07}$  \\
  $^{27}$Al &  $2.99\cdot 10^{-05}$ &  $^{64}$Zn &  $6.26\cdot 10^{-03}$ &    $^{99}$Ru &  $1.43\cdot 10^{-05}$ &   $^{145}$Nd &  $6.31\cdot 10^{-06}$ & $^{190}$Os &  $1.12\cdot 10^{-06}$  \\
  $^{28}$Si &  $4.95\cdot 10^{-02}$ &  $^{66}$Zn &  $3.98\cdot 10^{-02}$ &   $^{101}$Ru &  $4.42\cdot 10^{-05}$ &   $^{146}$Nd &  $3.28\cdot 10^{-06}$ & $^{192}$Os &  $1.45\cdot 10^{-06}$  \\
  $^{29}$Si &  $6.58\cdot 10^{-05}$ &  $^{67}$Zn &  $1.04\cdot 10^{-03}$ &   $^{102}$Ru &  $1.71\cdot 10^{-04}$ &   $^{148}$Nd &  $5.77\cdot 10^{-06}$ & $^{191}$Ir &  $5.36\cdot 10^{-07}$  \\
  $^{30}$Si &  $4.11\cdot 10^{-04}$ &  $^{68}$Zn &  $1.78\cdot 10^{-02}$ &   $^{104}$Ru &  $3.59\cdot 10^{-04}$ &   $^{150}$Nd &  $5.48\cdot 10^{-06}$ & $^{193}$Ir &  $1.81\cdot 10^{-06}$  \\
   $^{31}$P &  $1.34\cdot 10^{-04}$ &  $^{70}$Zn &  $5.67\cdot 10^{-03}$ &   $^{103}$Rh &  $4.19\cdot 10^{-04}$ &   $^{147}$Sm &  $6.73\cdot 10^{-06}$ & $^{194}$Pt &  $7.86\cdot 10^{-06}$  \\
   $^{32}$S &  $2.21\cdot 10^{-02}$ &  $^{69}$Ga &  $1.72\cdot 10^{-03}$ &   $^{105}$Pd &  $3.50\cdot 10^{-04}$ &   $^{149}$Sm &  $3.15\cdot 10^{-06}$ & $^{195}$Pt &  $1.56\cdot 10^{-05}$  \\
   $^{33}$S &  $8.62\cdot 10^{-05}$ &  $^{71}$Ga &  $6.98\cdot 10^{-04}$ &   $^{106}$Pd &  $1.77\cdot 10^{-04}$ &   $^{152}$Sm &  $7.49\cdot 10^{-06}$ & $^{196}$Pt &  $3.89\cdot 10^{-05}$  \\
   $^{34}$S &  $7.66\cdot 10^{-03}$ &  $^{70}$Ge &  $3.80\cdot 10^{-04}$ &   $^{108}$Pd &  $1.83\cdot 10^{-04}$ &   $^{154}$Sm &  $6.94\cdot 10^{-06}$ & $^{198}$Pt &  $1.02\cdot 10^{-04}$  \\
   $^{36}$S &  $2.19\cdot 10^{-07}$ &  $^{72}$Ge &  $3.56\cdot 10^{-03}$ &   $^{110}$Pd &  $2.45\cdot 10^{-04}$ &   $^{151}$Eu &  $8.28\cdot 10^{-06}$ & $^{197}$Au &  $2.72\cdot 10^{-05}$  \\
  $^{35}$Cl &  $1.37\cdot 10^{-04}$ &  $^{73}$Ge &  $1.70\cdot 10^{-03}$ &   $^{107}$Ag &  $2.49\cdot 10^{-04}$ &   $^{153}$Eu &  $5.07\cdot 10^{-06}$ & $^{199}$Hg &  $3.54\cdot 10^{-05}$  \\
  $^{37}$Cl &  $2.85\cdot 10^{-05}$ &  $^{74}$Ge &  $2.99\cdot 10^{-03}$ &   $^{109}$Ag &  $2.19\cdot 10^{-04}$ &   $^{155}$Gd &  $5.34\cdot 10^{-06}$ & $^{200}$Hg &  $7.01\cdot 10^{-06}$  \\
  $^{36}$Ar &  $4.28\cdot 10^{-03}$ &  $^{76}$Ge &  $6.00\cdot 10^{-03}$ &   $^{110}$Cd &  $1.03\cdot 10^{-10}$ &   $^{156}$Gd &  $8.01\cdot 10^{-06}$ & $^{201}$Hg &  $2.52\cdot 10^{-07}$  \\
  $^{38}$Ar &  $4.20\cdot 10^{-03}$ &  $^{75}$As &  $2.60\cdot 10^{-03}$ &   $^{111}$Cd &  $2.41\cdot 10^{-04}$ &   $^{157}$Gd &  $7.71\cdot 10^{-06}$ & $^{202}$Hg &  $1.75\cdot 10^{-08}$  \\
  $^{40}$Ar &  $1.13\cdot 10^{-07}$ &  $^{74}$Se &  $1.97\cdot 10^{-05}$ &   $^{112}$Cd &  $2.20\cdot 10^{-04}$ &   $^{158}$Gd &  $1.23\cdot 10^{-05}$ & $^{204}$Hg &  $2.31\cdot 10^{-09}$  \\
   $^{39}$K &  $1.13\cdot 10^{-04}$ &  $^{76}$Se &  $4.53\cdot 10^{-05}$ &   $^{113}$Cd &  $2.62\cdot 10^{-04}$ &   $^{160}$Gd &  $1.40\cdot 10^{-05}$ & $^{203}$Tl &  $3.23\cdot 10^{-09}$  \\
   $^{40}$K &  $2.19\cdot 10^{-08}$ &  $^{77}$Se &  $3.40\cdot 10^{-03}$ &   $^{114}$Cd &  $3.02\cdot 10^{-04}$ &   $^{159}$Tb &  $8.26\cdot 10^{-06}$ & $^{205}$Tl &  $1.72\cdot 10^{-10}$  \\
   $^{41}$K &  $5.64\cdot 10^{-06}$ &  $^{78}$Se &  $5.27\cdot 10^{-03}$ &   $^{116}$Cd &  $8.83\cdot 10^{-05}$ &   $^{161}$Dy &  $1.29\cdot 10^{-05}$ & $^{206}$Pb &  $1.32\cdot 10^{-08}$  \\
  $^{40}$Ca &  $3.50\cdot 10^{-03}$ &  $^{80}$Se &  $2.08\cdot 10^{-02}$ &   $^{115}$In &  $5.19\cdot 10^{-05}$ &   $^{162}$Dy &  $3.54\cdot 10^{-05}$ & $^{207}$Pb &  $1.12\cdot 10^{-08}$  \\
  $^{42}$Ca &  $6.87\cdot 10^{-04}$ &  $^{82}$Se &  $3.05\cdot 10^{-02}$ &   $^{117}$Sn &  $1.21\cdot 10^{-04}$ &   $^{163}$Dy &  $3.90\cdot 10^{-05}$ & $^{208}$Pb &  $1.18\cdot 10^{-08}$  \\
  $^{43}$Ca &  $9.00\cdot 10^{-06}$ &  $^{79}$Br &  $7.80\cdot 10^{-03}$ &   $^{118}$Sn &  $7.49\cdot 10^{-05}$ &   $^{164}$Dy &  $4.87\cdot 10^{-05}$ & $^{209}$Bi &  $8.00\cdot 10^{-09}$  \\
  $^{44}$Ca &  $5.73\cdot 10^{-05}$ &  $^{81}$Br &  $2.40\cdot 10^{-02}$ &   $^{119}$Sn &  $4.70\cdot 10^{-05}$ &   $^{165}$Ho &  $2.08\cdot 10^{-05}$ & $^{232}$Th &  $3.57\cdot 10^{-09}$  \\
  $^{46}$Ca &  $4.87\cdot 10^{-05}$ &  $^{78}$Kr &  $1.91\cdot 10^{-06}$ &   $^{120}$Sn &  $5.17\cdot 10^{-05}$ &   $^{166}$Er &  $3.47\cdot 10^{-05}$ &  $^{235}$U &  $8.07\cdot 10^{-10}$  \\
  $^{45}$Sc &  $6.78\cdot 10^{-06}$ &  $^{80}$Kr &  $5.56\cdot 10^{-06}$ &   $^{122}$Sn &  $3.21\cdot 10^{-05}$ &   $^{167}$Er &  $3.72\cdot 10^{-05}$ &  $^{238}$U &  $9.01\cdot 10^{-10}$  \\
  $^{46}$Ti &  $2.44\cdot 10^{-05}$ &  $^{82}$Kr &  $7.64\cdot 10^{-06}$ &   $^{124}$Sn &  $2.59\cdot 10^{-05}$ &   $^{168}$Er &  $2.23\cdot 10^{-05}$ & \\
  $^{47}$Ti &  $5.68\cdot 10^{-05}$ &  $^{83}$Kr &  $1.19\cdot 10^{-02}$ &   $^{121}$Sb &  $4.93\cdot 10^{-05}$ &   $^{170}$Er &  $1.27\cdot 10^{-05}$ & \\
  $^{48}$Ti &  $2.71\cdot 10^{-02}$ &  $^{84}$Kr &  $8.11\cdot 10^{-03}$ &   $^{123}$Sb &  $3.23\cdot 10^{-05}$ &   $^{169}$Tm &  $8.59\cdot 10^{-06}$ & \\
  $^{49}$Ti &  $1.31\cdot 10^{-03}$ &  $^{86}$Kr &  $2.40\cdot 10^{-03}$ &   $^{125}$Te &  $8.69\cdot 10^{-05}$ &   $^{171}$Yb &  $9.76\cdot 10^{-06}$ & \\
  \hline
  \multicolumn{10}{l}{The table is also published in machine-readable format.}
\end{tabular}

\end{table*}

\bsp	
\label{lastpage}
\end{document}